\documentclass[10pt,aps,twocolumn,pre,floatfix,superscriptaddress]{revtex4-1}
\usepackage{graphicx}
\usepackage{epstopdf}
\usepackage{amssymb,amsfonts,amsmath}
\usepackage{color}
\usepackage{float}

\newcommand{\bee}{\begin{equation}}
\newcommand{\ene}{\end{equation}}
\newcommand{\beea}{\begin{eqnarray}}
\newcommand{\enea}{\end{eqnarray}}

\newcommand{\pd}[2]{\frac{\partial #1}{\partial #2}}

\newcommand{\schmchn}{%
\begin{figure}[htbp]
  \begin{center}
   \includegraphics[width=3in,clip]{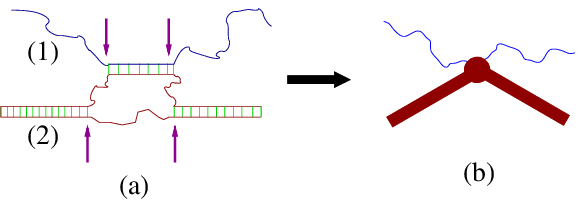}
 \end{center}
 \caption {(Color online) Schematic diagram of a strand exchange and
   the equivalent coarse-grained three-chain interaction $g_3$. (a) A
   single strand [blue line marked (1)] pairs with one strand of a
   bubble on a duplex [brown lines marked (2)].  The short vertical (green)
   lines indicate base pairings, with the energy per unit length
   $\epsilon$.  The junction weight $g_2$ is associated with
   each fork or the interface on a duplex.  In (a) there are four
   interfaces as indicated by the arrows.  (b) A coarse-grained
   version of (a) where the duplex is represented by a thick line
   interacting with the single line.  The filled circle  represents the
   three-chain interaction $g_3$.  }\label{fig:1}
\end{figure} }
\newcommand{\bdblock}{
\begin{figure}[htbp]
  \begin{center}
   \includegraphics[clip]{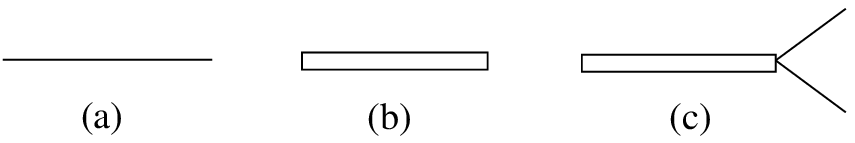}
 \end{center}
 \caption{Basic building blocks. Panel (a) represents $Z({\bf k},s)$
   for a Gaussian chain, (b) $Z_{\sf b}({\bf k},s)$ for a two chain
   bound state,(c) a Y fork representing the interface between a bound
   pair and two open strands. It has a weight $g_2$.}
   \label{fig:2}
\end{figure} }
\newcommand{\bdblockk}{
\begin{figure}[hb]
  \begin{center}
    \includegraphics[clip]{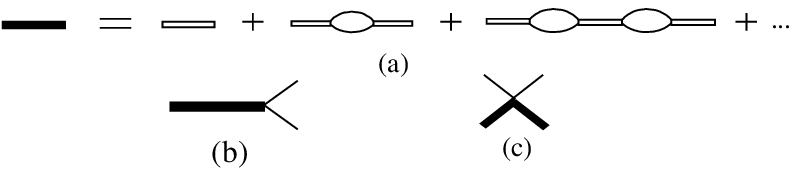}
 \end{center}
   \caption{ (a) The duplex partition function as an infinite series
     of bound pairs and bubbles, (b) Y fork for a duplex. (c) A
     three-chain interaction, 
$g_3$, involving a free chain and a duplex.
   }\label{fig:3}
\end{figure} }
\newcommand{\nobubble}{
\begin{figure}[htbp]
  \begin{center}
   \includegraphics[clip]{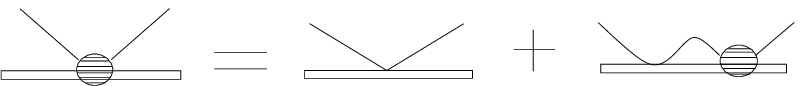}
 \end{center}
   \caption{ Interaction of free chain with a bound state in absence
     of bubbles. }\label{wall}
\end{figure} }
\newcommand{\thchint}{
\begin{figure*}[htbp]
  \begin{center}
   \includegraphics[width=6.5in,clip]{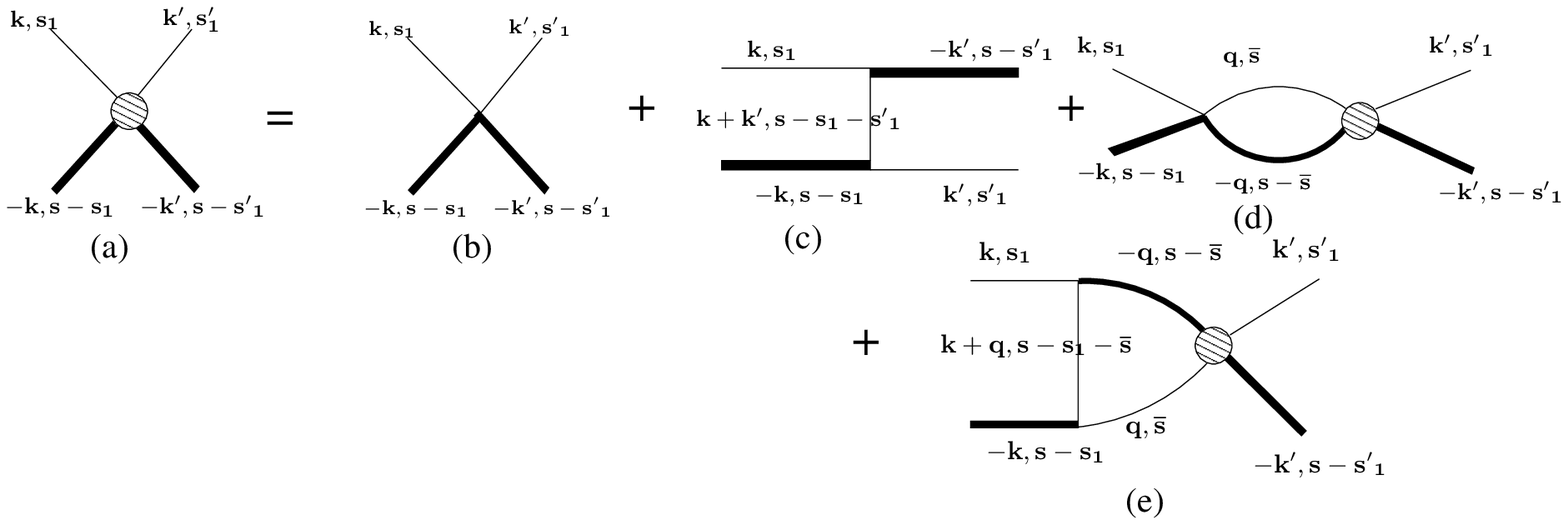}
 \end{center}
   \caption{Diagrammatic representation of the three-chain partition
     function.  The hatched circle is the effective interaction $W$. 
     This figure translates into an integral equation involving
     interactions to all order. }\label{fig:4}
\end{figure*}}
\newcommand{\plotH}{
\begin{figure}[t]
  \begin{center}
   \includegraphics[clip]{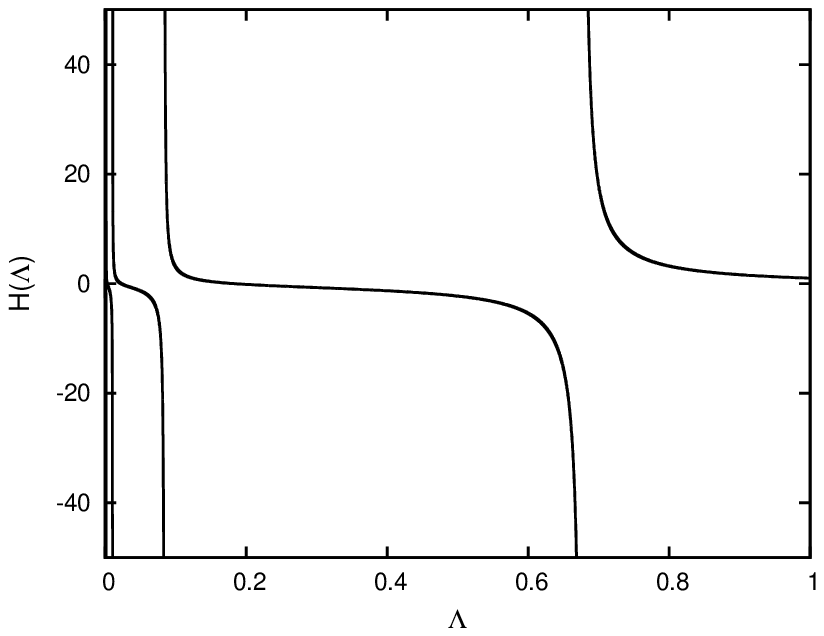}
 \end{center}
 \caption{Plot of $H$ as a function of $\Lambda$ showing zeros and
   divergences with $\Lambda_{*}=1$}\label{H-plot}
\end{figure}}
\newcommand{\comH}{
\begin{figure}[htbp]
  \begin{center}
   \includegraphics[width=\linewidth]{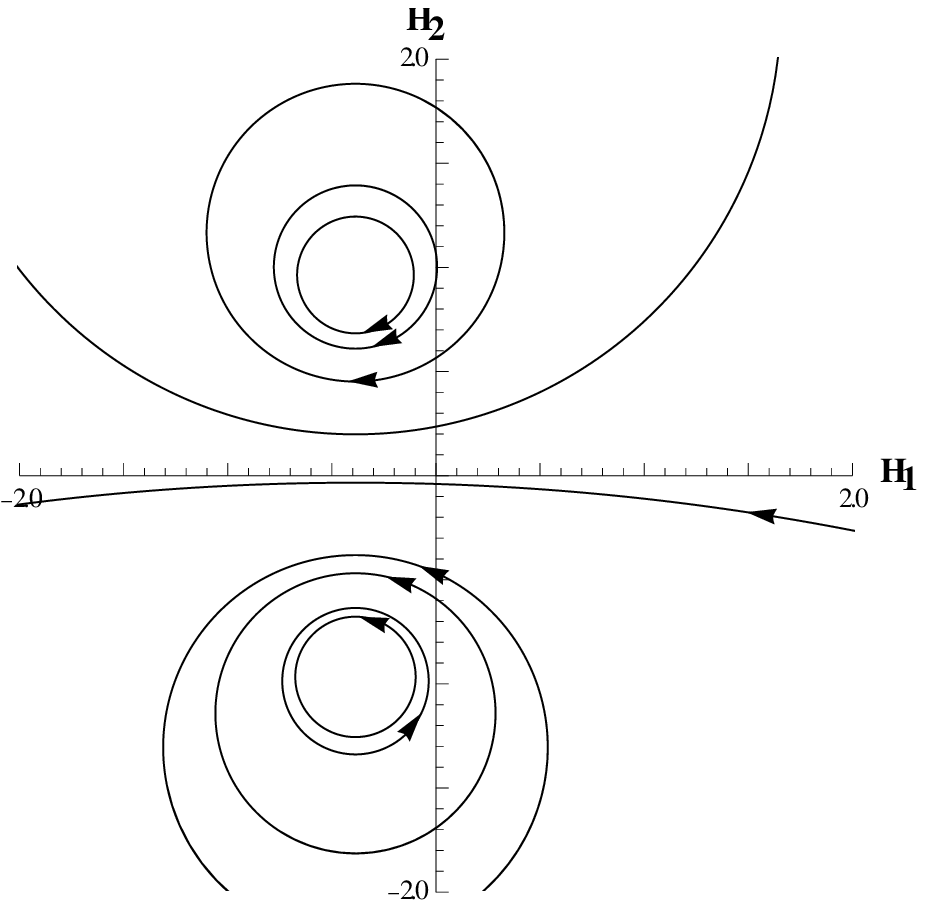}
 \end{center}
 \caption{Closed elliptical trajectories in the complex
   $H$ plane as $\Lambda$ is varied.  These are drawn for different
   starting values of $(H_1,H_2)$.  All loops in the upper half plane
   have the fixed point $(1+is_0)/(1-is_0)$ as a focus while
   $(1-is_0)/(1+is_0)$ as a focus for the lower half plane.
 }\label{complexH}
\end{figure}}
\newcommand{\laplace}{
\begin{figure}[htbp]
  \begin{center}
   \includegraphics[clip]{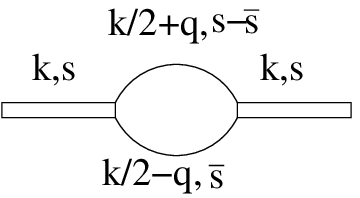}
 \end{center}
   \caption{ A bound state with one bubble showing variables in the
     Fourier-Laplace space variable obeying the {\bf k} and the
     $s$ conservations. }\label{laplace}
\end{figure} }
\newcommand{\gtwothree}{
\begin{figure}[htbp]
  \begin{center}
   \includegraphics[width=3in,clip]{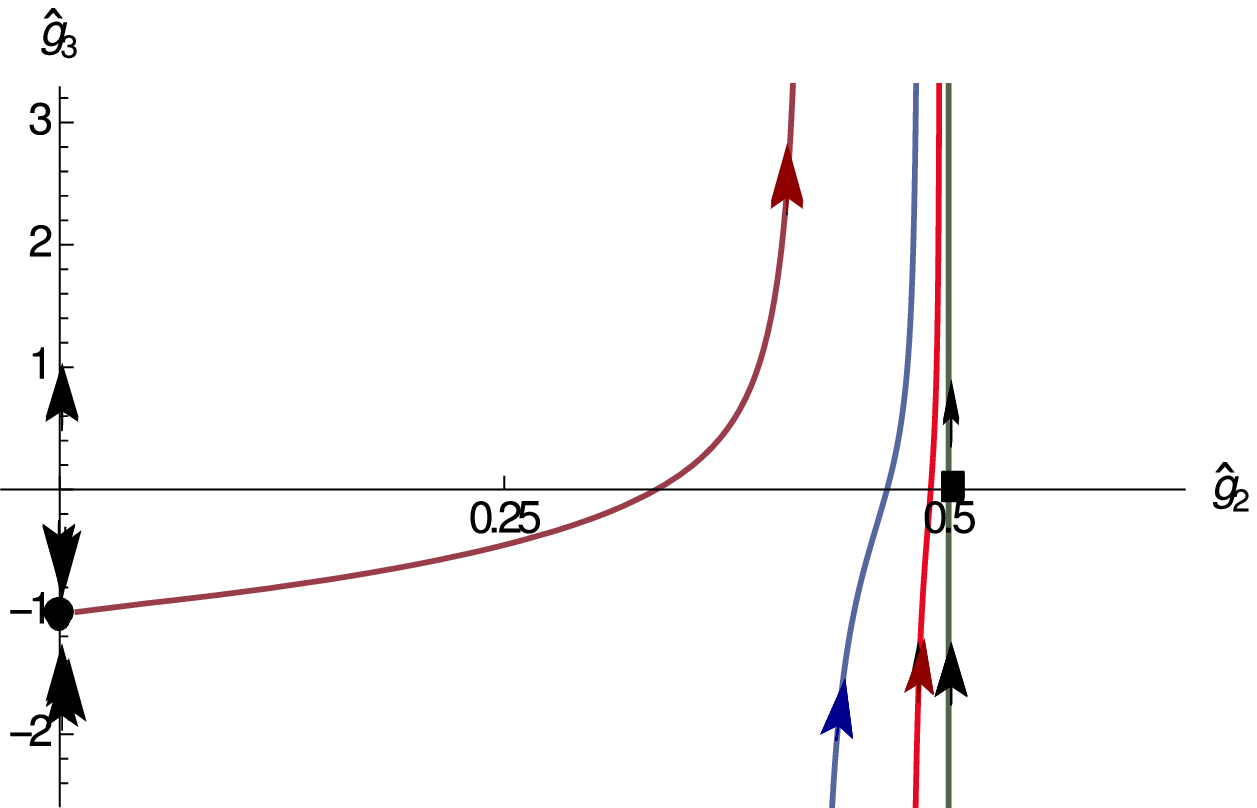}
 \end{center}
   \caption{ (Color online) Schematic flow diagram in the $\hat{g}_2$-$\hat{g}_3$
     plane.  $\hat{g}_2=0$ corresponds to the bound state (no bubbles)
     while $\hat{g}_2=\hat{g}_2^*$ (box) is the duplex melting point.
     The flow along the thick vertical line through
     $\hat{g}_2=\hat{g}_2^*$ is periodic.  Along the $\hat{g}_2=0$
     line, there is a stable fixed point $\hat{g}_3=\hat{g}_{3c}$
     (filled disk) which represents the peeling of one polymer from
     the rigid bound state.  A typical flow from a point away from the
     melting point is shown.  }
\label{fig:10}
\end{figure} }
\begin{document}
\title{Efimov-like phase of a three-stranded DNA  and the
  renormalization-group limit cycle}

\author{Tanmoy Pal}
\affiliation{Institute of Physics, Bhubaneswar 751005 India}
\email[email: ]{tanmoyp\string@iopb.res.in} 
\author{Poulomi Sadhukhan}
\affiliation{Institute of Physics, Bhubaneswar 751005 India}
\affiliation{Present address: Institut f\"ur Theoretische Physik,
  Universit\"at G\"ottingen, Friedrich-Hund-Platz 1, 37077
  G\"ottingen, Germany}
\author{Somendra M. Bhattacharjee} 
\affiliation{Department of Physics,
  Ramakrishna Mission Vivekananda University, Belur Math, 
  West Bengal 711202, India} 
\email[email: ]{somen\string@iopb.res.in; somen\string@rkmvu.ac.in}
\affiliation{Institute of Physics, Bhubaneswar 751005 India}

\begin{abstract}
  A three-stranded DNA with short range base pairings only is known to
  exhibit a classical analog of the quantum Efimov effect, viz., a
  three-chain bound state at the two chain melting point where no two
  are bound.  By using a non-perturbative renormalization-group method
  for a rigid duplex DNA and a flexible third strand, with base
  pairings and strand exchange, we show that the Efimov-DNA is
  associated with a limit cycle type behavior of the flow of an
  effective three-chain interaction.  The analysis also shows that
  thermally generated bubbles play an essential role in producing the
  effect.  A toy model for the flow equations shows the limit cycle in
  an extended three-dimensional parameter space of the two-chain
  coupling and a complex three-chain interaction.
  \end{abstract}

\maketitle

\section{Introduction}
As the storehouse of the genetic code, DNA is one of the most
important molecules in biology \cite{watson}. Structurally it is a
polymer made of bases, normally of four kinds, A, T, G, and C, whose
sequence along the chain codes for the amino acids of proteins.  DNA
is generally found in a double helical (dsDNA) form with two strands
attached to each other by the classic Watson-Crick type of hydrogen
bonding of A with T, and G with C; still many other alternative
conformations are known  \cite{dnaref}.  One such example is the triplex
or triple stranded DNA which can also be of several types.  In one
form, for special sequences, the third strand binds to a dsDNA via the
Hoogsteen base pairing and forms a triple helix
structure \cite{frank,djpatel,hoogs}. In another case, one dsDNA locally
melts forming a bubble and one strand of the bubble may pair with a
third strand, provided it has a sequence of matching complementary
bases.  This is called strand exchange \cite{strand}.  It was argued in
Refs.  \cite{jm,jayam,jmSG,pal} that near the thermal melting of a
dsDNA, the formation of large bubbles enhances the possibility of a
strand exchange involving each strand of the bubble.  This leads to an
effective long range attraction of the original pair, mediated by the
third strand.  As a result, a three-strand bound state is formed where
no two are bound.  Such a novel state of DNA, produced by
fluctuations, has been called an Efimov-DNA in analogy with the Efimov
effect in three-body quantum
mechanics \cite{efimov1,efimov2,efimov3,xx,brateen}.

DNA is nothing but two polymers, each one made of monomers (bases)
connected linearly, and with interaction between two monomers of the
two polymers only if they have the same contour length measured from
one physical end. This interaction is called the native DNA
interaction, and it produces the bound dsDNA. Thermal fluctuations can
break hydrogen bonds locally, making bubbles in the bound state.  When
all the base pairs are broken, either by thermal fluctuations or by a
force, releasing the two chains, one gets a melting or an unzipping
transition of DNA \cite{melt,unzip1,unzip2,unzip3}.  These transitions
are of importance in biology for their inherent functional utility,
and in polymer physics, as examples of ``few-chain'' problems
involving the interplay of polymer correlations and mutual
interactions (as opposed to polymer solutions \cite{degennes}).  To
this list of  few-chain  problems, is now added the
fluctuation-driven Efimov-DNA.

The Efimov effect was studied originally as a three-body quantum
mechanics problem by solving the Schr\"odinger equation in the Fadeev
approach \cite{efimov1,efimov2,efimov3,xx,brateen}.  Several
approximate methods were also used, most notable among which is the
use of the Born-Oppenheimer approximation for a separable short range
potential \cite{fonseca}.  Such calculations show the emergence of the
scale-free $1/r^2$ attraction at the quantum critical point of
unbinding, as does the polymer scaling of Ref.  \cite{jm}.  Field
theoretic methods were initiated much later on as an effective field
theory for few-body quantum mechanics.  Some successes were met in the
diatom approach where the effect was studied in the scattering of a
single particle from a diatom \cite{brateen,bedaque}.  The major effort
in these attempts was to see the Efimov effect as a universal effect,
emerging from a diverging length scale, here the scattering length.
It led to the idea of a limit cycle renormalization group (RG) flow \cite{ueda},
which was introduced in a different context in Ref.   \cite{glazek}.
The characteristic signature of the effect is the geometric sequence
($E_n = a^n E_0$) of energy eigenvalues, the Efimov tower, and it is
believed to emanate from this limit cycle behaviour.  It occurs only
at the point where the length scale diverges.  For nearby points one
still gets three-body bound states but with a finite number of states.
Although proposed in nuclear physics, experimental signatures for the
Efimov effect started pouring in only after the technological
developments in handling cold atoms \cite{zinner,jona,ultcol}.

The possibility of an Efimov-DNA was first pointed out by using a
scaling argument and by a real space renormalization group approach to
three polymers that can be implemented exactly for hierarchical
lattices \cite{jm,jayam}.  That the melting transition with a large or
diverging length scale is crucial (equivalent to infinite scattering
length in the quantum version) was clearly brought out by the polymer
scaling, that reproduces the $1/r^2$ interaction, where $r$ is the
distance between two polymers.  This interaction owes its origin to
the long range polymer correlations in a big bubble.  The importance
of the transition was also made clear in studies of the DNA problem in
lower dimensional fractal lattices.  In fact, a  mixed phase different
from the Efimov dna was
predicted, for which a quantum analog is not known \cite{jmSG}.  On the
polymer front, the strangeness of the long-range interaction is
evident from the renormalization-group analysis of two polymers with
$-g/r^{2}$ interaction in the presence of a short range
attraction \cite{sm1,sm2,kolom}. The unbinding transition is described,
as usual, by a fixed point, but that is not all.  First, the fixed
point is $g$-dependent, and, second, the order of the transition is
determined by the reunion exponent of the bubbles at the $g$-dependent
fixed point describing the unbound phase \cite{comm3}.  More unusual is
the possibility of complex fixed points.  This happens for $g>1/4$.
The complex fixed points are responsible for certain periodicity of
various thermodynamic quantities and is similar to the origin of the
Efimov tower.  A direct proof of the long range interaction is still
not possible but the emergence of a limit cycle behaviour in Euclidean
three dimensions was reported in Ref.  \cite{pal}.

The similarity between the zero temperature quantum problem and the
classical thermal system of polymers actually follows from an
imaginary time transformation of the quantum problem in the path
integral approach \cite{jm}.  For example, a path integral computation
of the quantum problem could identify polymer like
phases \cite{krauth}.  The fluctuations in the size of the polymer
bubbles near the melting point of dsDNA play a similar role as that
of quantum fluctuations near the unbinding transition of a pair of
particles.  The DNA bubbles correspond to the paths in the classically
forbidden region of the short range potential \cite{psent}.  In this
paper, we elaborate on the link between the Efimov-DNA and a limit
cycle behaviour in a renormalization group approach.  Our results show
that there are many other similarities of the results for a DNA with
the diatom trick used in the quantum version and it reinforces the
idea that the features of the quantum Efimov effect could be observed
in a classical setting of DNA in a solution.

The idea of RG is to look at a problem based on length scales and not
on the numerical values of the parameters per se.  How the various
parameters change with the length scale then tells us the behaviour of
the macroscopic system in the large size limit.  The procedure is to
(i) integrate out the small length scale fluctuations, especially
bubbles and (ii) then by rescaling generate a similar system but with
renormalized parameters.  The flows of the parameters as the scale is
changed give us all the crucial large length-scale results.  The
interpretation of the RG flows adopted here, discussed in detail
below, is different from the one generally done for
polymers \cite{smbjj,sm1,sm2} but similar in spirit as done in quantum
field theories, especially in the context of the Efimov
effect \cite{brateen}.  In general, an RG approach is expected to lead
to fixed points and separatrices, at most lines of fixed points.  The
fixed points represent states of the system which show scale
invariance under a continuous rescaling of lengths.  As pointed out
above, it is rather rare to see a limit-cycle-like behaviour because
its periodicity would produce a discrete scale invariance only.

\schmchn

Polymer problems traditionally start with a random walk or a Gaussian
polymer as the primary representation of a polymer. Most of the DNA
melting theories are of this type where the free model represents the
unbound states.  In the Gaussian polymer model (no self-interaction),
the melting transition is continuous \cite{comm0,comm00} so that the
closer one is to the melting temperature, the larger is the size of
the thermally generated bubbles.  In this approach the two-chain and
the three-chain problems have critical dimensionality $d=2$ and $d=1$
respectively \cite{smbjj}.  Evidently a strand exchange for a small
bubble in a three-chain system would, on scales larger than the bubble
size, look like a three-strand interaction (Fig. \ref{fig:1}).
Therefore, in three dimensions, one needs to consider irrelevant
variables and, so, a straightforward perturbative RG fails here. This
makes a complete analysis of the three-chain problem formidable. To
circumvent this, a different approach is adopted here.  Instead of
flexible Gaussian or semiflexible polymer representations, we model
the dsDNA as a sequence of rigid rod like bound segments and bubbles
made of flexible Gaussian chains at finite temperature. Thus the
melting point is approached from the bound state side via the
formation of bubbles.  We allow strand exchange in the bubble region
of a pair and study the behaviour of the three-body interaction
generated as the short distance cutoff is taken to zero.  In the
Fourier space, the limit corresponds to the upper cutoff
$\Lambda\to\infty$.  The resulting RG flow equation shows a limit
cycle behaviour, due to critical fluctuations, via nonphysical
complex fixed points.  Unlike the cases with special long range
forces, here all the interactions are strictly short ranged (like
hydrogen bonds for DNA).

The effect of fluctuations is not just restricted to the melting point
itself.  Even above the melting point, the bound state persists,
eventually melting at a temperature higher than the duplex melting
temperature.  The limit cycle that occurs at the melting point is
actually unstable as we move away from this special point.  The number
of turns in a way determines the number of bound states.  Beyond a
certain point all such states vanish.  This point or temperature will
be the melting point of the Efimov DNA.

The outline of the paper is given below.  The model is defined in Sec.
\ref{sec:model} in real space.  It involves the bound state of dsDNA
as a rigid chain and the third strand as a flexible chain.  The
interfacial term that helps in bubble formation through forking, and
the three chain interaction are defined here too.  A brief overview of
what is expected in a limit cycle RG is given in Sec.
\ref{sec:qual-descr}.  The calculations are done in the
Fourier-Laplace space.  The necessary rules for diagrams and the form
of the partition functions are given in Sec.
\ref{sec:diagr-defin-rules}.  We approach from the bound to the
unbound side.  For the two-chain bound state at finite temperatures,
forking is allowed with some energy cost. Two successive forkings
result in the formation of a bubble which can be infinite in number.
The corresponding duplex partition function and the melting of the
rigid DNA are discussed in Sec. \ref{sec:two-chain}.  The three-chain
case where we consider the duplex--free-chain interaction can be found
in Sec. \ref{sec:three-chain}.  Two cases are considered there.  A
case with no bubbles in the dsDNA is in Sec. \ref{sec:no-bubbles}
while the whole three-chain part is analyzed at the two body critical
point in Sec. \ref{sec:with-bubbl}.  A few details can be found in the
appendixes.  In particular, Appendix \ref{sec:limit-cycle-simple} is about
a toy example that extrapolates between the flow equations for the
three chain interaction with no bubbles and the same at the critical
melting point.  We end with a discussion of the results and their
experimental consequences in Sec. \ref{sec:discussion} and a short
summary in Sec.  \ref{sec:conclusion}.

\section{Model and result}\label{sec:model-result}
\subsection{Model}
\label{sec:model}
Our model of three polymer chains is defined, {\it \`a la}
Poland-Scheraga, through their partition functions.  Every monomer
lives in a $d$-dimensional space.  All chains are of equal length $N$.
They are tied at one end at origin in space while the other end may
also be tied together at a point ${\bf r}$, though the latter
constraint may be relaxed.  One end needs to be kept together to
prevent any chain from flying away, thereby facilitating the
bookkeeping for entropy.  The monomers on different chains interact
only when they are at the same space ($\bf r$) and length ($z$)
coordinates.  Such an attractive interaction corresponds to the native
base pairing of DNA.

The basic constituents of the model are the partition functions for a
single chain, for a bound pair, the weight $g_2$ for dissociation of a
pair or joining of two chains (Y forks), and the interaction $g_3$
between a single chain and a bound pair.  The two important parameters
in this problem are $g_2$ and $g_3$.

We first define the basic partition functions, viz., ({i}) ${\sf
  Z}({\bf r},N)$ for a single chain, and ({ii}) ${\sf Z}_b({\bf
  r},N)$ for a pure bound state of two strands.

\subsubsection{Single chain}
\label{sec:single-chain}
A single strand is a flexible Gaussian chain with
\begin{equation}
  \label{eq:9}
  {\sf Z}({\bf r},N)={\mu^{N\Lambda^{2}}}\ \frac{1}{(2\pi N)^{d/2}}\ e^{-\frac{{\bf  r^{2}}}{2N}},
\end{equation}
where $\mu^{N\Lambda^{2}}$ is the total number of configurations, and
$\Lambda^{-1}$ is a short distance or microscopic cutoff.  For a
Gaussian chain the overall size $R$ of a polymer scales as
\begin{equation}
  \label{eq:5}
 R^2\sim N,
\end{equation}
which allows us to set the dimension of $N$ as
\begin{equation}
  \label{eq:6}
  [N]={\rm L}^2, \quad {\rm when\ } [r]=[\Lambda^{-1}]={\rm L},
\end{equation}
the square bracket $[...]$ indicating the dimensionality of the
enclosed entity.  In Eq. \eqref{eq:9}, $\Lambda$ is used to make $N$
dimensionless in the $\mu$-dependent factor.  The unconstrained
entropy of a free chain is taken as $\propto \ln \mu$ per unit length
to avoid the problem of an infinite entropy of a continuous chain.
The Gaussian factor in Eq. \eqref{eq:9} is the probability density of
finding the end of the polymer at ${\bf r}$, so that the total
partition function after integration over all space is dimensionless.

\subsubsection{Bound state, Y fork, duplex, and  $g_2$}
\label{sec:bound-state-y}
The second partition function needed is for the two-chain bound state
which is taken as a rigid rod with $\epsilon\Lambda^{2}$ as the
binding energy per unit length. The bound DNA with one end fixed can
rotate in space as a whole but can not bend. The partition function of
the bound state of length $N$ is given by
\begin{equation} 
\label{eq:21}
{\sf Z}_{\sf b}({\bf r},N)=\frac{1}{4\pi}\; e^{-\epsilon N\Lambda^{2}}
                         \;\delta({\bf r}-N\Lambda\hat{\bf n}),  
\end{equation}
where $\hat{\bf n}$ is a unit vector  giving the direction of the
rigid rod.   

Now, there are finite temperature fluctuations in the form of pair
breaking.  The bound state then locally dissociates into two single
strands to form a Y fork.  The two free strands may rejoin to produce
a bubble.  We assign an interface weight $g_2$ in the partition
function for every Y fork.  This is like an extra interfacial
contribution and is referred to as the ``co-operativity factor''
 \cite{jmSG}.  It is shown below Eq. \eqref{eq:22}, that dimensionwise
\begin{subequations}
\begin{eqnarray}
\label{eq:12}
[g_{2}^{2}]&=&[\Lambda]^{4-d},\\
                &=&[\Lambda]^{1},\quad (d=3).\label{eq:23}
\end{eqnarray}
\end{subequations}
The importance of bubbles is well-recognized both in the biological
functions and in physical properties of DNA.  There have been many
studies, in recent times, both theoretical and experimental, on the
nature and functions of the bubbles under various situations, like DNA
under a force or topological
constraints \cite{son,king,jeon08,strick,jeon10}, in breathing
dynamics \cite{altan,ambjor,alexand,Sicard}, in
hysteresis \cite{hyst,kumhyst}, with semiflexibility \cite{jeon06}, etc.

The introduction of finite temperature bubbles makes the bound state
flexible and as a result it can bend.  We name this bound state with
bubbles a {\it duplex}.  The $g_2$-dependent partition function ${\sf
  Z}_{\sf d}$ is discussed in Sec \ref{sec:two-chain}.  We see there
that the formation of large bubbles, by thermal fluctuations leads to
the melting of a duplex into two free chains, at a critical value of
$g_2=g_{2c}$.  The role of $g_2$ may be represented by the sequence of
partition functions
\begin{equation}
  \label{eq:18}
  \underbrace{\sf Z_b}_{\{g_2=0\}} \stackrel{{\rm  crossover}}{\longrightarrow} 
 \underbrace{\sf Z_d}_{\{g_2<g_{2c}\}} 
 \stackrel{{\rm melting}}{\longrightarrow}  
 \underbrace{\sf Z^2}_{\{g_2>g_{2c}\}},
\end{equation}
where the change from a rigid to an elastic one by $g_2$ is a
crossover (not a transition).

\subsubsection{Strand exchange and  $g_3$}
\label{sec:strand-exchange-g_3}
For the three-polymer system, we consider the situation where a pair
is in the bound or duplex state and the other one is free.  The third
chain is allowed to interact with a free strand of a bubble allowing
it to form a duplex locally.  This is called strand exchange.  Figure
\ref{fig:1} illustrates this process schematically.

The pair interaction together with the strand exchange would, in
principle, be sufficient to formulate the three-chain DNA problem, but
in a renormalization group approach, the three-chain bound state is
described by a three-chain interaction.  Therefore, in anticipation of
its generation, we allow a three-chain interaction (between a single
chain and a duplex) $g_3$.  This interaction parameter is our
three-body coupling.  We see below Eq. \eqref{eq:24} that it has the
dimensionality
\begin{subequations}
\begin{eqnarray}
\label{eq:25}
[g_{3}]&=&[\Lambda]^{2-d},\\
          &=&[\Lambda]^{-1}\quad (d=3).
\end{eqnarray}
\end{subequations}
The dimensionless three-chain parameter may now be constructed
as \cite{commg3}
\begin{eqnarray}
  \label{eq:19}
 H(\Lambda)&=&-\; \frac{g_{3}}{4g_{2}^{2}}\Lambda^{2} ,
\end{eqnarray}
with a factor of $4$ for convenience.  

The partition function for three chains depends on both $g_2$ and
$g_3$ but for a duplex it depends only on $g_2$.

\subsection{Qualitative description}\label{sec:qual-descr} 
Our aim is to see the effect of strand exchange on the three-chain
system near the duplex melting point where large sized bubbles are
expected.  Right at the melting point, we may concentrate on how $g_3$
or $H$ evolves as $\Lambda\to\infty$ for large $N$.  A flow of $H$
from zero to $+\infty$ is an indication of a three-chain bound state
[note the negative sign in Eq. (\ref{eq:19})], the Efimov-DNA case of a
bound three-chain system where no two are bound.

In the RG approach the flows of parameters are obtained in a few
steps.  With a reciprocal space upper cutoff $\Lambda$ (a short
distance scale $\sim \Lambda^{-1}$), the effects over a range
$\Lambda$ to $\Lambda-d\Lambda$ are taken into account by redefining
the problem for scales up to $\Lambda-d\Lambda$.  A subsequent
rescaling brings back the problem to the original scale with
renormalized parameters.  The changes in the parameters, as continuous
variables, gives us the flow equations or $\beta$ functions
\begin{equation}
  \label{eq:20}
  \Lambda\frac{\partial H}{\partial \Lambda}=\beta(H).  
\end{equation}
The behaviour of a system is then characterized by the flows which
generally terminate at stable fixed points (or at infinity) separated
by unstable fixed points. The fixed points represent the phases and
the phase transitions in the system.  This is the generic picture of
RG and this is where the three-chain problem stands out.

In our approach, $g_2$ is the control parameter for the duplex
melting.  We first determine the critical point of melting by locating
the critical value $g_{2c}$ at which a suitably defined length scale
$\xi$ diverges.  At this particular point we determine the RG flow or
the $\beta$ function for $H(\Lambda)$.  Quantitatively this is
implemented by calculating the third virial coefficients, obtained
from the connected three-chain partition function (i.e., for polymers
connected by the interactions).

A naive use of the definition, Eq. \eqref{eq:19}, suggests the form
\begin{subequations}
\begin{equation}
  \label{eq:35}
\beta(H)=2H,  \quad ({\rm{naive}}).
\end{equation}
This however gets modified by the effects of strand exchange and other
fluctuations to a form
\begin{equation}
  \label{eq:36}
\beta(H)= 2H+{\cal F}(H), \quad ({\rm with\ renormalization}),
\end{equation}
\end{subequations}
and all the nontriviality comes from these additional terms.  In
general, it has a form
\begin{equation}
  \label{eq:26}
\beta(H)= -AH^2 +B H +C, 
\end{equation}
$A, B, C$ being all real.  In conventional cases, $B$ is mainly
determined by the naive dimensional analysis, while a nonzero $A$ is
the extra addition of length rescaling and renormalization.  A
constant term $C$ is unusual and appears if there are marginal
parameters, which do not change with length scales.  An example of a
marginal parameter is $g$ of the inverse square interaction mentioned
in the Introduction.  This identification may be turned around to
argue that a constant term in the $\beta$ function signals the presence
of some, may be hidden, marginal parameter in the problem.

If $C$ in Eq. \eqref{eq:26} is such that there are two real roots of
$\beta(H)=0$, the standard picture remains valid with one stable and
one unstable fixed points, but not so if the roots are complex
conjugate pairs [for $C<-B^2/(4A)$].  Such is the case for the problem
in hand.  We find
\begin{equation}
   \label{eq:4}
   \Lambda\frac{\partial H}{\partial \Lambda}=-A(H-H_0)(H-H_0^*),
\end{equation}
where $H_0,H_0^*$ form a complex-conjugate pair.  

The procedure we adopt to derive Eq.~\eqref{eq:4} is different from the
conventional RG way.  The traditional approach is to take into account
the effects at the short distance level to redefine the parameters on
a larger length scale. Instead of such an approach we determine the
effective parameter as an integral equation and then use a thin-shell
integration method to get the $\Lambda$ dependence of $H$ by demanding
the existence of a cutoff independent limit.  From this we reconstruct
the $\beta$ function.  Then we argue that at the duplex melting point the
$\beta$ function of the dimensionless scaled three-body interaction
parameter $H$ has the same form as that of Eq.~\eqref{eq:4} and the
limit cycle describes the three-body bound Efimov states.

The nonconformity with the standard picture of fixed points has a far
reaching consequence of converting the continuous scaling symmetry at
the unstable fixed point to a discrete symmetry.  The continuous
scaling symmetry at a real fixed point leads to power law behaviours
of physical quantities.  Contrary to that, complex fixed points invoke
a limit-cycle-type behaviour in the RG flow trajectories.  An outcome
of the generated periodicity is a discrete scaling symmetry and the
relevant parameter, here $H$, repeats itself in a log periodic manner.
In the quantum language, this discrete symmetry leads to the Efimov
tower of the energies.

\subsection{Diagrammatic definitions and rules}\label{sec:diagr-defin-rules} 

Since we shall be using diagrammatic representations for many
equations, it is prudent to define our model in terms of diagrams and
the rules for computations.  To take advantage of the convolution
property of the Fourier and the Laplace transforms, we work in the
Fourier-Laplace space $({\bf k}, s)$ instead of $({\bf r},N)$.  The
conventions for these transforms are
\begin{eqnarray}
\label{eq:7}
{\hat Z}({\bf k},N)&=&\int {\sf Z}({\bf r},N)e^{-i{\bf k.r}}\; d^dr,\\
Z({\bf k},s)&=&\int_{0}^{\infty}{\hat Z}({\bf k},N)e^{-Ns}dN,
\end{eqnarray}
with the inverse  transforms defined as
\begin{eqnarray}
  \label{eq:8}
{\sf Z}({\bf r},N)=\frac{1}{(2\pi)^d}\int {\hat Z}({\bf k},N)e^{i{\bf
    k.r}} d^dk,\\
Z({\bf k},N)=\frac{1}{2\pi i}\oint Z({\bf k},s)e^{Ns}ds.  \label{eq:1}
\end{eqnarray}
The contour for the integration in Eq. \eqref{eq:1} is the usual
Mellin's contour.
\bdblock

The dimensionalities of the partition functions, as per our
conventions, are as follows
\begin{subequations}
\begin{eqnarray}
  \label{eq:2}
&&[{\sf Z}({\bf r},N)] = {\rm L}^{-d}, [{\hat Z}({\bf k},N)]={\rm L}^0,\\
&& [ Z({\bf k},s) ]={\rm L}^{2}, {\rm \ with\ }  [s]={\rm L}^{-2}. \label{eq:3} 
\end{eqnarray}
\end{subequations}
These $L$ dependencies are used to identify the dimensionalities of the
remaining parameters.

\subsubsection{Free chain}
In the Fourier-Laplace space the single chain partition function, Eq.
\eqref{eq:9}, becomes
\begin{equation}
  \label{eq:10}
Z({\bf k},s)=\frac{1}{s-\Lambda^{2}\ln\mu+\frac{k^{2}}{2}}.
\end{equation}
To be noted here is  that the free energy per unit length comes from the
pole of Eq. \eqref{eq:10} in the complex-$s$ plane.  This partition
function, Eq. \eqref{eq:10}, to be called a propagator, is represented
by a solid line in Fig. \ref{fig:2}(a).

\subsubsection{Bound state}
The rigid bound state of Eq. \eqref{eq:21} is direction dependent.
For simplicity we take an average over all directions by integrating
over the solid angle subtended by $\hat{\bf n} $ (see Appendix
\ref{appen_bound_state}).  The corresponding Fourier-Laplace
transformed partition function is given by
\begin{equation}
  \label{eq:11}
  Z_{\sf b}({\bf k},s)=\frac{1}{k\Lambda}\arctan\frac{k\Lambda}
  {s+\epsilon\Lambda^{2}}\stackrel{k\rightarrow0}{=}\frac{1}{s+\epsilon\Lambda^{2}}. 
\end{equation}
The $k\rightarrow0$ form of $Z_{\sf b}({\bf k},s)$ can be used close
to the melting defined shortly.  Here also the pole in the complex
$s$ plane gives the free energy of the rigid bound state.  The
bound-state partition function of Eq. \eqref{eq:11} is represented by
an unfilled rectangular box in Fig. \ref{fig:2}(b).

At finite temperatures, the inclusion of the Y forks gives the duplex
partition function $Z_{\sf d}({\bf k},s)$ represented by a filled
black box in Fig.  \ref{fig:3}(a).

The Y-fork junction, $g_2$, is represented in Fig. \ref{fig:2}(c) by a
vertex where a rectangular box (a bound pair) and two solid lines
(free chains) meet.  The diagram in Fig. \ref{fig:3}(c) represents an
interaction between a duplex and a free chain. The interaction is the
three-body coupling constant given by $g_{3}$.

\bdblockk

\subsection{${\bf k}$ and $s$  Conservation}
As all the partition functions have translation invariance we have
${\bf k}$ conservation at each vertex.  Following the standard
nomenclature, the ${\bf k}$ vectors are called ``momentum."   So
momentum is conserved at each vertex.  Also cutoff $\Lambda$ is
called a momentum cutoff.

The Y fork and the three-chain interaction can take place anywhere
along the length of the polymers which are infinitely long.  This
invariance (translational invariance along the contour) leads to an
$s$-conservation at any point.  In other words, at a junction
$s$-values get distributed, but the total remains the same.  More
details are discussed in Appendix \ref{appen_rules}.

\section{Two Chains}\label{sec:two-chain}
At first let us find the duplex partition function.  In the grand
canonical ensemble (Laplace space) the singularity of the partition
function closest to the origin gives the free energy of the system;
contributions from others are suppressed in the thermodynamic limit.
Whenever there is a switching of the nearest singularity due to a
change in some parameter of the system, we have a phase transition.
Henceforth all the calculations are done in $d=3$.

Considering an arbitrary number of bubbles we can write the finite
temperature bound state as a sum of an infinite number of diagrams
shown in Fig. \ref{fig:3}(a).  In terms of the Laplace variable $s$,
the duplex partition function, denoted by $Z_{\sf d}({\bf k},s),$ can
be written as a geometric series
\begin{eqnarray}
  \label{eq:13}
Z_{\sf d}({\bf k},s)&=&Z_{\sf b}({\bf{k}},s) +g_{2}^{2}\;Z_{\sf b}({\bf{k}},s)\; I_0 
\;Z_{\sf b}({\bf{k}},s)+....\nonumber\\
          &=&\frac{1}{\frac{1}{Z_{\sf b}({\bf{k}},s)}-g_{2}^{2}I_{0}},\label{eq:22}
\end{eqnarray}
where $I_{0}$, the single bubble contribution, is given by
\begin{eqnarray}
  \label{eq:14}
  I_{0}&=&\int\frac{d{\bf q}}{(2\pi)^{3}}\; \frac{d\bar{s}}{2\pi i}  
  \; Z\left(\frac{{\bf k}}{2}-{\bf q},\bar{s}\right)\;  Z\left(\frac{{\bf
        k}}{2}+{\bf q},s-\bar{s}\right)\nonumber\\ 
  &=&\frac{1}{2\pi^{2}}\left[\Lambda-\sqrt{s'+k^{2}/4}\;\arctan\frac{\Lambda} 
    {\sqrt{s'+k^{2}/4}}\right],
\end{eqnarray}
and $s'=s-2\Lambda^{2}\ln\mu$.  Equation \eqref{eq:22}, with Eqs.
\eqref{eq:2} and \eqref{eq:3}, sets the dimension of $g_2$ as quoted
in Eq.  \eqref{eq:12}.

To evaluate $I_{0}$ we do the $\bar{s}$ integral by the method of
residues.  See Appendix \ref{appen_rules} for details.  The only
contribution comes from the simple pole at
$\bar{s}=\Lambda^{2}\ln\mu-({\bf k}/2-{\bf q})^{2}/2$.  In the limit
$(s'+k^{2}/4)\rightarrow0$ and $\Lambda$ finite, which is the relevant
limit near the transition point, we have 
\bee
I_{0}=\frac{1}{2\pi^{2}}\left[\Lambda-\frac{\pi}{2}\sqrt{s'+k^{2}/4}\right].
\ene So the duplex partition function becomes \bee Z_{d}({\bf
  k},s)=\left\{s+\epsilon\Lambda^{2}- \frac{1}{2\pi^{2}}
  g_{2}^{2}\left[\Lambda-\frac{\pi}{2}
    \sqrt{s'+k^{2}/4}\right]\right\}^{-1}.\label{dressed} 
\ene
We identify here three different singularities in the partition
function $Z_{\sf d}$, which correspond to three distinct states.  The branch
point singularity of Eq. \eqref{dressed} at $s=2\Lambda^{2}\ln\mu$,
owing its origin to $Z$, gives the completely unbound (denatured)
state.  This is the high temperature phase. The singularity at
$s=-\epsilon\Lambda^{2}$ corresponds to the completely bound state
when $g_2=0$.  This, being the singularity of $Z_{\sf b}$, is the zero
temperature phase and does not survive when $g_2\neq 0$. The third
singularity $s'_{*}$ comes from the zero of the denominator of
$Z_{\sf d}(0,s)$. As $s'_{*}$ continuously evolves with $g_2$ from
$s=-\epsilon\Lambda^{2}$, it corresponds to a bound state with
bubbles. In the absence of $g_{2}$, i.e., in absence of any interface
or junction point, there can be two states only; the system stays
either in the completely bound state or in the completely unbound
state.  There could be a denaturation transition, necessarily first
order, by changing $\epsilon$ or $\mu$.  This case is of no interest
to us.  The presence of the interface alters the nature of the bound
state because of the bubbles and also makes the transition critical
[see Eq. \ref{eq:18}].

Our main aim is to concentrate on the behaviour of the system near
duplex melting where the contributions from the bubbles (loops) of
large sizes dominate the duplex partition function.  
In the small
$s'$ limit with  $k=0$, $s'_{*}$ is given by
\begin{equation}
  \sqrt{s'_{*}}=-\frac{\Delta t}{2\pi^{2} g_{2}^{2} \Lambda^{-2}} , \label{eq:15} 
\end{equation}
where $\Delta t\equiv (2\pi)^{3}(2\ln\mu+\epsilon) -4\pi
g_2^2\Lambda^{-1}$.  $s'_{*}$ can be identified as the difference of
free energies between duplex and two free chain states. So, $\Delta t$
is a measure of deviation from the duplex melting point. The equation
$\Delta t=0$ \cite{deltat} gives the critical point as
\begin{equation}
  \label{eq:47}
  g_{2c}=\sqrt{2 \pi^2 (2 \ln\mu + \epsilon)\Lambda}.
\end{equation}

The thermal melting of a dsDNA can also be illustrated from this
model \cite{fisher}. Using Eq. (\ref{eq:15}), we define a diverging length scale
$\xi$ in the following way
\begin{equation}
    \label{eq:16}
    s'_*\sim\xi^{-2}, \ {\rm with}\ \xi\sim |\Delta t|^{-1}.
\end{equation}%
If we now make a scale change such that, for arbitrary $b$, ${\bf
  k}\rightarrow b^{-1}{\bf k}$, the length scale changes as
$\xi\rightarrow b\xi$. And as $s'_{*}$ is the free energy difference,
the free energy scales as $f\rightarrow b^{-2}f$. This is the
continuous scale invariance satisfied at the thermal dsDNA melting.

By tuning $g_{2}$ the length scale $\xi$ can be made divergent for
some critical value of $g_{2}$, $g_{2c}$. Beyond it, the system goes
to a stable high temperature phase with two free chains. The full
duplex partition function can be written for small $s'$ as
\begin{equation} 
  Z_{\sf d}({\bf k},s)=\frac{(2\pi)^{3}}
  {{2\pi^{2}g_{2}^{2}\left[-{\xi}^{-1}+\sqrt{s'+
          k^{2}/4}\right]}},\label{duplex}
\end{equation}
which explicitly shows the $\xi$ dependence.

\section{Three Chains}\label{sec:three-chain}
Now consider the three-chain problem. The effect of thermal
fluctuations (bubbles) are very important in our model. To make this
more clear, we first consider the case where the bubbles are not
allowed.  Next, we consider the case with bubbles.

\subsection{No bubbles: $g_2=0$}
\label{sec:no-bubbles}
Let us first consider the case of a free chain interacting with a
bound state. Here we set $g_2=0$ such that there are no bubbles in the
bound state. This problem with interaction up to all orders can be
solved by solving the diagrammatic integral equation shown in Fig.
\ref{wall}  (see Appendix \ref{appen_rules}).
\nobubble

The bare single chain bound state contact interaction is $g_{3}$ and
we denote the corresponding renormalized interaction by $V$.
Evaluations of Fig. \ref{wall} in $d-$dimensions give us
\begin{equation}
  V=-g_{3}-g_{3}V\int \frac{d{\bf q}}{(2\pi)^{d}}\;
  \frac{d\bar{s}}{2\pi i}\; Z({\bf q},\bar{s}) \; Z_{\sf b}({\bf -q},s-\bar{s}).\label{eq:24}
\end{equation}
Here $V$ is taken as a function of $\Lambda$ but not ${\bf q}, s$.
The above equation can be used to obtain the dimension of $g_3$ as
quoted in Eq. \eqref{eq:25}.  The $\bar{s}$ integral is evaluated by
the method of residues as
\begin{equation}
  V=-g_{3}-g_{3}V\int_{0}^{\Lambda}\frac{\Omega_{d}}{(2\pi)^{d}}\frac{q^{d-1}dq}{s-\Lambda^{2}\ln\mu
    +\Lambda^{2}\epsilon+q^{2}/2},
\end{equation}
where $\Omega_{d}$ is the surface area of the $d$-dimensional unit
hyper sphere.  The unbound phase consists of two independent members,
a rigid bound pair and a free polymer, with the system free energy
determined by \mbox{$s=\Lambda^{2}(\epsilon-\ln\mu)$}.  Considering the
system to be in this unbound state we get
\begin{equation}
 \label{eq:27a}
 \hat{V}=-\hat{g_{3}}-\hat{g_{3}}\frac{\hat{V}}{d-2},
\end{equation}
where 
\begin{equation}
  \label{eq:27}
 \hat{g_{3}}=\frac{2\Omega_{d}\Lambda^{d-2}}{(2\pi)^{d}}g_{3},
\quad \hat{V}=\frac{2\Omega_{d}\Lambda^{d-2}}{(2\pi)^{d}}V, 
\end{equation}
are dimensionless quantities.  Equation  \eqref{eq:27a} can be rewritten as
\begin{equation}
 \label{eq:27b}
 \hat{g}_3= - \frac{\hat{V}}{1+\frac{\hat{V}}{d-2}}.
\end{equation}
The RG flow equation of $\hat{g}_3$ is obtained simply by
differentiating with respect to $\Lambda$, keeping $V$ constant.  The
result is
\begin{equation}
\label{eq:41}
  \Lambda\frac{\partial\hat{g}_3}{\partial\Lambda}=(d-2)\hat{g}_3+\hat{g}_3^{2},
\end{equation}
where the linear term on the right hand side can be linked to
dimensional analysis, Eq. \eqref{eq:27}. The quadratic term is the
loop contribution.  A small loop, quadratic in $g_3$, on a bigger
scale would look like an effective interaction, modifying the coupling
constant.

Writing down the RG flow equation in terms of $\Lambda$ for the bare
values is similar to the use in quantum problems.  Here one studies
the flow of the bare values for a fixed renormalized coupling, while
the converse is done in the usual polymer RG.  Consequently the
stability of the fixed points are of opposite nature compared to the
polymer RG fixed points of say Ref.  \cite{smbjj,sm1,sm2}.  The limit
of $\Lambda\to\infty$ with say $\hat{g}_3$ constant corresponds to the
limit of an infinitely strong potential but of shrinking width,
approaching a $\delta$ function.  Compared to this depth of the potential
the binding energy $V$ is very small, i.e. $V/g_3\to 0$.  In this
situation as the range is taken to be zero, the bound state looks like
it is close to the threshold.  In the terms of bubbles, the length-scale for
the bubbles, be it above or below the transition ($V\lessgtr 0$),
looks much larger compared to the range of the interaction and,
therefore, closer to the critical point which has a diverging length
scale.  This explains why the critical point in this scheme
corresponds to a stable fixed point.

The flow equation is similar to the RG flow of the two-chain
coupling \cite{smbjj} with a stable and an unstable fixed point.  For
$d=3$, the stable fixed point $\hat{V}^*=-1$ corresponds to the
critical point of unbinding and $\hat{V}^*=0$ represents the unbound
state.  This is expected because the bound DNA acts as a single rigid
polymer with no internal structure so that the problem is effectively
like the unbinding of two dissimilar DNA strands.

\subsection{With bubbles:  $g_2\neq 0$}
\label{sec:with-bubbl}

Now consider the three-chain case allowing thermal-fluctuation
generated bubbles in the bound state.  Here we always consider
situations where any two of the three chains have formed a duplex and
the other free chain is interacting with that duplex. This
consideration simplifies the problem immensely.   We formulate
our analysis at  the two-chain melting point $g_2=g_{2c}$ to find the
three-chain partition
function.   From this partition function the effective three-body coupling at the duplex
melting point can be determined.  There are no small parameters in the
problem and therefore we need to sum terms up to infinite order or
equivalently solve the integral equation shown diagrammatically in Fig
\ref{fig:4}.

Let us generalize the effective interaction $V$ of Sec.
\ref{sec:no-bubbles} to a three-chain vertex function as $W$ (see
Appendix \ref{appen_rules}), which in general depends on the input and
the output momenta and the $s$ values [Fig. \ref{fig:4}(a)].  Two
successive Y forks producing a strand exchange at a small separation
would look like a three-chain interaction (see Fig. \ref{fig:4}(c)).
This is an ${\rm O}(g_2^2)$ term.  One may also couple this
strand-exchanged configuration to the rest of the three-chain
interactions, Fig. \ref{fig:4}(e), generating a term of ${\rm O}(g_2^2
W)$.  The $g_3$-dependent terms of Fig. \ref{wall} also occur but with
the replacement of the bound propagator (unfilled rectangles) by that
of the duplex (filled rectangles), Fig. \ref{fig:4}(b) and
\ref{fig:4}(d).  By combining all these, we have,
\begin{widetext}
\begin{eqnarray}
  \label{eq:17}
  W({\bf k,k'},s_{1},s_{1}',s) &=&2g_{2}^{2}Z({\bf k+k'},s-s_{1}-s_{1}')-g_{3}\nonumber\\
  &&\ \ +2g_{2}^{2}\int \frac{d{\bf q}}{(2\pi)^{3}}\frac{d\bar{s}}
  {2\pi i}Z({\bf q},\bar{s})Z({\bf k+q},s-s_{1}-\bar{s})Z_{\sf d}(-{\bf q},s-\bar{s})
  W({\bf q,k'},\bar{s},s_{1}',s)\nonumber\\
  &&\ \ -g_{3}\int \frac{d{\bf q}}{(2\pi)^{3}}\frac{d\bar{s}}
  {2\pi i}Z({\bf q},\bar{s})Z_{\sf d}(-{\bf q},s-\bar{s})W({\bf q,k'},\bar{s},s_{1}',s).
\end{eqnarray}
Notice the factor of $2$ in the diagrams with strand exchange because
the chains are distinguishable \cite{comm1}.

If we do the $\bar{s}$ integration by residues, the only contribution
is from the pole of $Z({\bf q},\bar{s})$ at
$\bar{s}=\Lambda^{2}\ln\mu-q^{2}/2$.  This relation between $\bar{s}$
and ${\bf q}$ is analogous to the real space relation for size [Eq.
\eqref{eq:6}], which means the free chain is in a relaxed
state \cite{comm2}.  Small distortions around the average size of a
free polymer in equilibrium can be described by the Gaussian
distribution around its average.  Therefore this residue guarantees
that no special large stretching takes place in a strand-exchange and
the free chain remains more or less like an average chain.  So we have
\begin{eqnarray}
  W({\bf k,k'},s_{1},s_{1}',s)&=&2 g_{2}^{2}\;  Z({\bf
    k+k'},s-s_{1}-s_{1}')  -  g_{3}\nonumber\\ 
  && +  \int\frac{d^3q}{(2\pi)^{3}}\;  \left(\frac{2g_{2}^{2}}
    {s-s_{1}-2\Lambda^{2}\ln\mu+q^{2}/2+(k^{2}+q^{2})/2+{\bf
        k.q}}  -  g_{3}\right)\nonumber\\ 
  &&\hspace{2cm}\times  W({\bf q,k'},s_{1}',s)\; Z_{\sf d}(-{\bf
    q},s-\Lambda^{2}\ln\mu+{\bf q}^{2}/2).\label{eq:28}
\end{eqnarray}
To simplify let us do the angle averaging, i.e. replacing $Z_{\sf d}$
by Eq. \eqref{duplex}, so that $W, Z_{\sf d}$ are functions of the
magnitudes of the wave vectors.  The remaining angular integral from $
{\bf{ k.q}}$ can be done.  Assuming the external single chains are in
their relaxed states such that $s_{1}=\Lambda^{2}\ln\mu-k^{2}/2$ and
$s_{1}'=\Lambda^{2}\ln\mu-k'^{2}/2$ we have the angle averaged
partition function
\thchint
\begin{eqnarray}
  W(k,k')&=&\frac{g_{2}^{2}}{kk'}\ln\frac{s''+k^{2}+k^{'2}+kk'}{s''+k^{2}+k^{'2}-kk'}-g_{3} \nonumber\\  
  &&+4\pi\int_{0}^{\Lambda}\frac{dq}{(2\pi)^{3}}
  q^{2}\left(\frac{g_{2}^{2}}{qk}\ln\frac{s''+q^{2}+k^{2}+qk} 
    {s''+q^{2}+k^{2}-qk}-g_{3}\right)W(q,k')Z_{\sf d}(-q,s+q^{2}/2),
  \label{eq:29} 
\end{eqnarray}
where $s''=s-3\Lambda^{2}\ln\mu$. 

\subsubsection{Critical case: $g_2=g_{2c}$}
\label{sec:crit-case:-g_2=g_2c}

The limit required here is $s''\rightarrow0$ which asserts that all
chains are critical simultaneously as we have three chains now.  To
make sure that we are around the two-body critical point we take the
limit $\xi\rightarrow\infty$ in $Z_{d}(-q,s+q^{2}/2),$ Eq.
\eqref{eq:16}. In this limit we expect to have significant
contributions from the loop diagrams, and so we neglect the tree
diagrams.  By defining dimensionless quantities, like $H$ in Eq.
\eqref{eq:19},
\begin{equation}
  \label{eq:30}
 \overline{W}(q,k') = q W(q,k'), 
\end{equation}
and using Eq. (\ref{duplex}), at the melting point,  we have
\begin{equation}
  \overline{W}(k,k')=\frac{8}{\sqrt{3}\pi}\int_{0}^{\Lambda}\frac{dq}{q}
  \left[\ln\frac{q^{2}+kq+k^{2}}{q^{2}-kq+k^{2}}+2kq\frac{H(\Lambda)}{\Lambda^{2}}\right]  
  \overline{W}(q,k').\label{master}
\end{equation}
The main reason behind the difference in the form of Eqs.
\eqref{master} and \eqref{eq:24} lies in the criticality of the
duplex partition function used here.  Since it is possible to consider
the $k'\to 0$ limit of Eq. \eqref{master}, the functional dependence
of $W$ on $k'$ is not important for our calculations.  We therefore
suppress $k'$ hereafter.

\subsubsection{Scale-free limit}
\label{sec:scale-free-limit}
In the limit $H\rightarrow 0$ and $\Lambda\rightarrow\infty$ there is
no scale left in the problem because $g_2$ has already been tuned to
its critical value where $\xi \to\infty$.  In this scale-free limit,
the eigen-function type equation for $\overline{W}$ is
\begin{equation}
  \label{eq:32}
  \overline{W}(k)={\cal I}_{k,q} \overline{W}(q)\equiv
  \frac{8}{\sqrt{3}\pi}\int_{0}^{\infty}\frac{dq}{q} 
  \left[\ln\frac{q^{2}+kq+k^{2}}{q^{2}-kq+k^{2}}\right]  
  \overline{W}(q).
\end{equation}
Since $\overline{W}$ is dimensionless, a manifestly dimensionless form
of Eq. \eqref{eq:32} is obtained by replacing $k,q$ by
$\hat{k}=k/\Lambda_*, \hat{q}=q/\Lambda_*$, for some arbitrary
$\Lambda_*$.  Furthermore there is a large-$k$ -- small -$k$ duality of
integral operator ${\cal I}_{\hat{k},\hat{q}}$ which suggests two
degenerate solutions for Eq. \eqref{eq:32}.  This is a consequence of
the invariance of the integral operator under a transformation
\begin{equation}
  \label{eq:33}
  \hat{Q}=\frac{\Lambda_*}{q}, \hat{K}=\frac{\Lambda_*}{k},  \ {\rm
    with }\quad {\cal
    I}_{\hat{k},\hat{q}}\equiv {\cal I}_{\hat{K},\hat{Q}},
\end{equation}
Thus if $f(k/\Lambda_*)$ is an eigenfunction of ${\cal I}_{k,q}$, then
so is $f(\Lambda_*/k)$.  The general solution of $\overline{W}(k)$ can
then be written as a sum of the two degenerate solutions.

Taking note of the scale-free form, we can have a power law ansatz
\begin{equation}
  \label{eq:31}
 \overline{W}(k)\approx \left( \frac{k}{\Lambda_*}\right)^{s},
\end{equation}
which on substitution in Eq.  \eqref{eq:32} yields
\begin{equation}
  \label{eq:34}
  s=\frac{16}{\sqrt{3}}\ \frac{\sin (\pi s/6)}{\cos (\pi s/2)}.
\end{equation}
This equation has solutions for pure imaginary values, $s=\pm is_{0}$,
with 
\begin{equation}
  \label{eq:48}
  s_{0}=1.5036,\quad {\rm or}\quad \exp(\pi/s_0)=8.0713...  ,
\end{equation}
which is different from $22.7$ obtained by Efimov.

The solution for $\overline{W}$ is a linear combination of $\exp[\pm i
s_0 \ln(k/\Lambda_*)]$, which can be recast in a trigonometric form
\begin{equation}
\label{wform}
\overline{W}(k)=C\cos\left(s_{0}\ln\frac{k}{\Lambda_{*}}\right), 
\end{equation}
with   $C, \Lambda_{*}$ as two arbitrary constants.

\subsubsection{For $\Lambda < \infty$}
\label{sec:lambda--infty}
In the general case, ($\Lambda<\infty$) we may still proceed to find
the $\Lambda$ dependence of $H(\Lambda)$ by assuming that
$\overline{W}$ approximately retains its form ( as in Eq.
(\ref{wform})) by changing only its constants \cite{bedaque}.

Defining the function $f(\Lambda)=\frac{H(\Lambda)}{\Lambda^{2}}$ we
can rewrite Eq.  (\ref{master}) as
\begin{equation}
  \overline{W}(k)=\frac{8}{\sqrt{3}\pi}\int_{0}^{\Lambda}\frac{dq}{q}
  \left[\ln\frac{q^{2}+kq+k^{2}}
    {q^{2}-kq+k^{2}}+2kqf(\Lambda)\right]\overline{W}(q).\label{remaster} 
\end{equation}

$\overline{W}$ is related to the third virial coefficient of the
system. For this $\overline{W}$ must be independent of $\Lambda$ which
is introduced arbitrarily. We take advantage of this fact to compare
the value of $\overline{W}$ for two infinitesimally different
$\Lambda$s. The cutoff independence is preserved by equating the
residual pieces to zero.  By integrating over a small shell of radius
$\Lambda dl$ we have
\begin{eqnarray}
  \overline{W}(k)&=&\frac{8}{\sqrt{3}\pi}\int_{0}^{\Lambda e^{-dl}}\frac{dq}{q}
  \left[\ln\frac{q^{2}+kq+k^{2}}{q^{2}-kq+k^{2}}+2kqf(\Lambda)\right]\overline{W}(q)\\
  &&+\frac{8}{\sqrt{3}\pi}\left[\ln\frac{\Lambda^{2}+k\Lambda+k^{2}}
    {\Lambda^{2}-k\Lambda+k^{2}}+2k\Lambda f(\Lambda)\right]\overline{W}(\Lambda)dl.
\end{eqnarray}
Rescaling back $\Lambda\rightarrow\Lambda e^{dl}$ and retaining terms
up to order $dl$  we have
\begin{eqnarray}
  \overline{W}(k)&=&\frac{8}{\sqrt{3}\pi}\int_{0}^{\Lambda}\frac{dq}{q}
  \left[\ln\frac{q^{2}+kq+k^{2}}{q^{2}-kq+k^{2}}+2kqf(\Lambda)\right]\overline{W}(q)\nonumber\\  
  &&+dl\frac{8}{\sqrt{3}\pi}\left(2k\pd{f(\Lambda)}{l}\int_{0}^{\Lambda}dq\overline{W}(q)+
    \left(\frac{2k}{\Lambda}+2f(\Lambda)k\Lambda\right)\overline{W}(\Lambda)\right).\label{step1}
\end{eqnarray}
In the previous step we used the approximation $k \ll\Lambda $ such
that
\begin{equation}
\ln\frac{\Lambda^{2}+k\Lambda+k^{2}}{\Lambda^{2}-k\Lambda+k^{2}}\approx\frac{2k}{\Lambda},
\hspace{0.5cm}({\rm for\ } k\ll \Lambda). 
\end{equation}
Now using Eq. ($\ref{remaster}$) in Eq. (\ref{step1}) we can easily
arrive at the differential equation
\begin{equation}
  \frac{1}{\Lambda}\left[\Lambda\pd{H}{\Lambda}-2H\right]\int_{0}^{\Lambda}dq
  \overline{W}(q)+[1+H]\overline{W}(\Lambda)=0,
\end{equation}
where $d\Lambda=\Lambda dl$.  This equation for $H$ is already of the
form of Eq. \eqref{eq:36}, except that the terms in addition to the
naive $2 H$ term is dependent on $\Lambda$.

In principle, a renormalization-group $\beta$ function is not expected to
have any explicit cutoff dependence.  The $\Lambda$-independent flow
equation is derived below from the full form for $H(\Lambda)$.  To do
so, by inserting $\overline{W}(x)=C\cos(s\ln x),$ where
$x=\frac{\Lambda} {\Lambda_{*}},$ in that equation we obtain,
\begin{equation}
  \pd{H(x)}{x}\frac{\sin(A+B)}{x}+\frac{H(x)}{x^{2}}(s_{0}-1)\sin(A+B)+ 
  \frac{s_{0}\cos(A-B)}{x^{2}} -\frac{\sin(A-B)}{x^{2}}=0,
\end{equation}
where $A=s_{0}\ln x$ and $B=\arctan(\frac{1}{s_{0}})$. We can express
the left hand side of the above equation as an exact differential,
\begin{equation}
  \pd{}{x}\left[\frac{H(x)}{x}\sin\left(s_{0}\ln x+\arctan\left(\frac{1}{s_{0}}\right)\right)+ 
    \frac{1}{x}\sin\left(s_{0}\ln x-\arctan\left(\frac{1}{s_{0}}\right)\right)\right]=0. 
\end{equation}
Setting the boundary condition such that the integration constant
vanishes,  we get
\begin{equation}
  H(\Lambda)=-\frac{\sin\left(s_{0}\ln
      \frac{\Lambda}{\Lambda_{*}}-\arctan\left(\frac{1}{s_{0}}\right)\right)}
  {\sin\left(s_{0}\ln
      \frac{\Lambda}{\Lambda_{*}}+\arctan\left(\frac{1}{s_{0}}\right)\right)}.\label{theh}
\end{equation}
Having found out the $\Lambda$ dependence of $H$ we can now derive its
RG flow equation by simply taking a derivative of Eq. ($\ref{theh}$).
The $\beta$ function of $H$ is given by
\begin{equation}
\label{betah}
\Lambda\pd{H}{\Lambda}=\beta(H)\equiv
2H -    \frac{1}{2} \; (1+s_0^2)\; (H+1)^2 ,
\end{equation}
\end{widetext} 
which is of the form of Eq. \eqref{eq:26}, with 
\begin{equation}
  \label{eq:38}
 A = -C = \frac{1}{2} \;(1+s_0^2), B=1-s_0^2 .
\end{equation}

The specialty of the flow equation, Eq. \eqref{betah}, is the emergence
of complex conjugate fixed points, $H_0, H_0^*$, with
\begin{equation}
  \label{eq:37}
  H_{0}=\frac{(1+is_{0})}{(1-is_{0})},
\end{equation}
so that one recovers the form of Eq. \eqref{eq:4}.  Further  consequences are
discussed below.

\subsubsection{Complex fixed points and periodicity}\label{sec:results-discussion}

Because of the complex fixed points the flow of $H$ consists of closed
trajectories in the complex $H$-plane.  This is at the duplex melting
point, a fixed point for $g_2$ in the renormalization group sense,
and, therefore, the flows remain planar in the complex $H$ plane.  We
define a new variable $\zeta=(H-H_{0})/(H-H^{*}_{0})$ which is nothing
but a conformal mapping of $H$.  The flow equation of $\zeta$ is the
equation of a unit circle
\begin{equation}
  \label{zetaeq}
  \frac{\Lambda}{\zeta}\pd{\zeta}{\Lambda}=2is_{0}.
\end{equation}

In critical phenomena normally one expects the occurrence of real
fixed points in the flow equation of relevant parameters.  In this
particular RG scheme, a stable fixed point serves as the critical
point for the corresponding parameter. In the vicinity of the critical
point the system is scale-free because the system length scale
diverges with exponent $\nu$ which is related to the difference of the
fixed points.  Unlike this, when we have a limit cycle, the continuous
scaling symmetry breaks down and the relevant parameter is log
periodic. A power law $f(x)\sim x^{-\nu}$ for real $\nu$, is converted
to an oscillatory form $f(x)\sim e^{-i|\nu|\ln|x|}$ where $\nu$ is
imaginary $(\nu=i|\nu|)$. This occurs because the difference between
the complex fixed points is a purely imaginary quantity now.  A
consequence of this is the log periodicity of $H$ with respect to
$\Lambda$ in Eq. \eqref{theh}.

There is another convenient way to visualize the closed trajectories.
Decompose $H$ into its real and complex parts by writing
$H=H_{1}+iH_{2}$ to get two interdependent differential equations:
\begin{subequations}
\begin{eqnarray}
  \Lambda\pd{H_{1}}{\Lambda}&=&-\frac{1}{2}\left[s_{0}^{2}
    \left\{(1+H_{1})^{2}-H_{2}^{2}\right\}\right.\nonumber\\
&&\hspace{1cm}+\left.(1-H_{1}^{2})-H_{2}^{2}\right],\label{eq:42}\\
  \Lambda\pd{H_{2}}{\Lambda}&=&(1-H_{1})H_{2}-s_{0}^{2}(1+H_{1})H_{2}.\label{eq:43}
\end{eqnarray}
\end{subequations}
By solving these two equations simultaneously for different initial
values we get closed elliptical trajectories. All ellipses in the
upper-half plane have one common focus at one complex fixed point
$H_{0}$ and the other foci are at different places.  A similar thing
happens in the lower-half plane too, with a common focus at
$H_{0}^{*}$. The trajectories which start on the real line ($H_{2}=0$)
always stay on the real line. These are shown in Fig.
\ref{complexH}.

These closed loops change over to the real form of Eq. \eqref{eq:41}
as $g_2$ is detuned from the critical point.  A toy model for this
smooth crossover is discussed in Appendix \ref{sec:limit-cycle-simple} that
shows the specialty of the closed loop in the three-dimensional space
of $g_2$ and complex $g_3$ (in dimensionless forms).
\comH

\subsubsection{Discrete scale invariance}
\label{sec:discr-scale-invar}

A simple inspection of Eq. (\ref{theh}) shows us that $H$ is log
periodic.  As already mentioned, at the duplex melting point $g_2$ has its
fixed point value and so, from the definition of $H$, the
dimensionless three-body interaction energy, $\hat{g}_3\sim
g_3\Lambda$ obeys a flow equation very similar to that of $H$.  If we
start from $\hat{g}_{3}=0$ we arrive at the negative infinity as
$\Lambda$ is increased.  At this point $\hat{g}_{3}$ jumps to positive
infinity and decreases to negative infinity again as $\Lambda$ is
increased further. This is shown in Fig.  \ref{H-plot}.
\plotH
This behaviour goes on and $\hat{g}_{3}$ runs into negative infinity
whenever the denominator of Eq.  (\ref{theh}) becomes zero.  This
occurs at the points
\begin{equation}
\label{efimovlambda}
\Lambda_{n}=\Lambda_{*}\left(e^{\frac{\pi}{s_{0}}}\right)^{n}\exp
\left[\frac{\arctan(s_{0})-\frac{\pi}{2}}{s_{0}}\right],  
\end{equation}
where $n$'s are integers.  We therefore see the emergence of a
discrete scale invariance in this three-chain problem even though the
melting itself or the three-chain interaction {\it per se} has no indication
of this sort.

As $\hat{g}_{3}$ can also be interpreted as the three-body binding
energy, at those values of $\Lambda$ we get the three-body Efimov
bound states in the quantum case. The corresponding energy spectrum of
the quantum three-particle system would follow a geometric relation
given by
\begin{equation}
  \label{eq:44}
 E_{n+1}/E_n=e^{-2\pi/s_0},
\end{equation}
where $E_n$ is the $n$th energy state.  So the energies of the Efimov
states are related by a factor of $e^{2\pi/s_{0}}$.  We observe from
Eq. (\ref{zetaeq}) that two successive windings around the unit circle
are also related through the factor $e^{\pi/s_{0}}$ . As $H$ has
nontrivial values at these points we conclude that every jump from
one energy level to the next one corresponds to one winding of $\zeta$
around the closed trajectory. It can also be observed that
$\hat{g}_{3}$ goes to zero at the points
\begin{equation}
\Lambda_{n}=\Lambda_{*}\left(e^{\frac{\pi}{s_{0}}}\right)^{n}\exp
\left[\frac{\frac{\pi}{2}-\arctan(s_{0})}{s_{0}}\right], 
\end{equation}
where the numerator of Eq.  $(\ref{theh})$ becomes zero. At these
special points we do not have to introduce three-body coupling. So, in
this picture every jump corresponds to a switching of one Efimov state to 
another one and it is associated with a complete winding around a limit
cycle. These states are crowded more and more as one goes in the
direction of zero energy. They are infinite in number.

To summarize, we see that each of $g_2$ and $g_3$, acting alone on its
own, allows critical points in the form of melting or dissociation,
well described by the conventional renormalization group fixed points.
These points show a continuous scale invariance; under a rescaling of
the system by any factor, $L\to bL$ for any $b$, a critical system
remains critical, statistically identical.  In contrast, at such a
fixed point for $g_2$, $g_3$ shows a cyclic behaviour, better
described as a ``limit cycle'' behaviour, in the complex plane,
because of the emergence of a periodicity.  The log periodicity
induces a discrete scale invariance, $L\to b_nL$ for a discrete set
$b_n$, breaking the continuous symmetry expected at the critical value
of $g_2$ for two chains.  In quantum mechanics, this leads to an
infinite set of energy eigenstates in a three-particle system at the
point where the energy of any pair should have been zero.  In the
context of DNA, at the melting point of a double stranded DNA where
the strands are not bound to each other, a third strand induces a
binding of a size much larger than the hydrogen bond length.  In other
words the infinite correlation length scale of a duplex DNA gets
transmuted to a finite value in the presence of a third one when each pair
is supposed to be critical.

\subsection{Off-critical: $g_2\neq g_{2c}$}
\label{sec:crit-g_2n-g_2c}
So far we have considered the case of the critical two-chain case.
For the general situation, $g_2\neq g_{2c}$, we need to go back to Eq.
\eqref{eq:29}, and we also need the flow equation for $g_2$. Instead, we
may take a heuristic approach.  In terms of the dimensionless two- and
three-body constants $\hat{g}_3, \hat{g}_2\sim g_2\Lambda^{-1/2}$,
with $ H\sim \hat{g}_3/\hat{g}_2^2$, the expected form of the RG flow
equation for $g_2$ is
\begin{equation}
  \label{eq:40}
  \Lambda \pd{\hat{g}_2}{\Lambda}=\beta_2(\hat{g}_2),
\end{equation}
with an unstable fixed point $\hat{g}_2=0$ for the bound phase and a
stable one at $\hat{g}_2=\hat{g}_c$ for the duplex melting.  At these
points $\beta_2(\hat{g}_2)=0$. With these, we may formally write
\begin{equation}
  \label{eq:39}
  \Lambda \pd{\hat{g}_3}{\Lambda} = \frac{2\hat{g}_3}{\hat{g}_2}\ \beta_2(\hat{g}_2)
  - 4 \; \hat{g}_2^2 \; \beta(H,\hat{g}_2),
\end{equation}
relating the $\beta$ function for $\hat{g}_3$ with the others.  As
expected, for the duplex melting point with $\beta_2(\hat{g}^*_2)=0$,
the flow of $\hat{g}_3$ is the same as that of $H$ and therefore
$\hat{g}_3$ is to be described by the pair of complex fixed points.
However, for the $\hat{g}_2=0$ fixed point, to get back Eq.
\eqref{eq:41}, $\beta(H)$ of Eq. \eqref{betah} is not sufficient
because it does not yield a $\hat{g}_2$-independent limit.  This
indicates that the contributions of the off-critical terms in $W$, Eq.
\eqref{eq:29}, are important.  This can be taken as a signal that the
periodicity that develops at the duplex melting point for $g_3$ or
$\hat{g}_3$ do not survive in the off-critical limit.  A flow diagram
is shown in Fig. \ref{fig:10} which depicts a few cycles before merging
with the flow at $g_2=g_{2c}$.

\gtwothree

\section{Discussion}
\label{sec:discussion}
At this point we would like to place the results of this paper in a
broader context.  We do so in three different contexts, namely (i) as
a DNA problem, (ii) as an Efimov effect, and (iii) more formally as a
renormalization-group problem.  Let us first discuss the last two
issues, as elaborated upon in the Introduction.  The results, in
conjunction with the previous works, provide a possible testing ground
for the quantum Efimov physics in a classical environment, namely the
melting of DNA which occurs at temperatures in the range 60--100C.
Here thermal fluctuations play the role of quantum fluctuations
\cite{comm4}.  It was shown earlier that the fluctuation induced long
range inverse-square attraction has a natural basis in the polymer
scaling.  Here we showed how the polymer phase transitions in the
various limiting situations allowed us to construct an extrapolation
formula for the renormalization group $\beta$ function that shows the
development of the limit cycle behaviour.  This is also important in
the general theory of renormalization group where examples of limit
cycles are rather few.

Let us now look at it as a DNA problem.  The paper builds on the model
of a stiff duplex \cite{pal} and extends the study of the third virial
coefficient of the three-polymer system to the region around the
duplex melting point (at temperature $T_c$).  This extension, which
goes beyond Ref.  \cite{pal},  shows that the fluctuation-induced
Efimov DNA is not just a specialty of the melting point but it also exists
over a region where the duplex should have been unbound.  Although the
fractal-like lattices of Ref.   \cite{jm,jayam,jmSG} showed the
possibility of the three-chain thermodynamic phase, the lack of a
metric or distance forbade any analysis of the inverse square
law attraction responsible for the Efimov effect.  This gap is now partly filled
by the analysis of this paper.  Short of a direct proof of the
inverse-square attraction, the limit cycle behaviour is similar to the
complex fixed points known in systems with such long range
interaction.

There are several issues which might be amenable to experimental
verification.  The third strand could be made of alternating short
sequences of both the strands.  (i) A direct test of the Efimov-effect
in DNA would be a measurement of the melting temperature $T_t$ of the
triple-chain system to see if $T_t >T_c$.  The melting is expected to
be first-order in nature.  The difficulty of course lies in separating
the melting of the Efimov DNA from that of a Watson-Crick and
Hoogsteen paired triple-stranded DNA.  (ii) A different thermodynamic
study would be the virial coefficients via the osmotic pressure of
dilute solutions \cite{virial}.  The third virial coefficient will give
$H(\Lambda)$, Eq.  (\ref{eq:19}).  Near the melting of duplex, the
cutoff parameter may be chosen as the bubble length scale $\xi$, Eq.
(\ref{eq:16}).  Therefore, for $\Lambda\sim \xi^{-1}\sim \sqrt{\Delta
  t}$, the virial coefficient is expected to show an oscillatory
behaviour whose periodicity is determined by the (nonuniversal) Efimov
number $s_0$.  (iii) More detailed information might be obtained from
small angle neutron scattering (SANS) or light scattering.  One of the
signatures of the fluctuation induced long range interaction is the
$1/k$ divergence of $W(k)$ in Eq.  (\ref{eq:30}), since $\overline{W}$
is bounded.  Such a divergence in the interaction is observable as a
zero-$k$ peak in the scattered intensity in SANS or light scattering
 \cite{Liuscatt}.  (iv) It is possible to design tailor-made
environments where one might test some of the details that have gone in
the theory.  For example, one may use the geometry of two similar
strands maintained at a distance larger than the hydrogen bond length
to prevent direct pairing and then allow a complementary strand to
form bonds via strand exchange (Fig.  \ref{fig:1}) with both the
strands.  The force needed to unzip one of the original strands can
then be measured to verify the inverse square nature.  Such
experiments will be similar in spirit to the measurement of the
fluctuation induced Casimir force or the entropic interaction in DNA
solution \cite{casimir,verma}.  (v) An assumption that has gone in the
theory, actually in most field theoretic calculations (see Ref.
 \cite{comm2}), is that the third chain between two contacts with the
other two strands is in a relaxed state.  This condition may be
verified by using a carefully constructed bubble and then placing a
third chain.  A labeled chain (with say heavier isotopes) will help
in separating out the scattering from this chain and provide
information on its configuration.  (vi) A different experiment would
be to study the DNA in a narrow pore.  The constraint prevents the
large loop formation, thereby cutting off the long range interaction.
In this situation, the Efimov-DNA formation is not likely to happen
but a novel finite size effect would be expected \cite{grosberg}.  The
effects would be observable in the configuration of the labeled chain
and even in thermodynamic quantities.  Such experiments of DNA in a
pore has been attempted but not at the level that may explore the
large loops near DNA melting \cite{Reisner}.  (vii) One may go beyond
the three-chain problem to the possibility of a four or more chain bound
state, or even a gel formation in a many strand solution through the
Efimov interaction.  In such a gel, just above the duplex melting
point, the mesh size of the network would be similar to the bubble
size.  A theoretical study of this gelation phenomenon and the elastic
properties of the gel remain an open and interesting topic.

\section{Conclusion}\label{sec:conclusion}
This paper presents a model of a three-stranded DNA as a three-chain
polymer system, which shows mathematically analogous results as that
of Efimov physics, namely, the possibility of a three-chain bound
state when no two are bound.  The existence of a three-stranded DNA
bound state (Efimov DNA) at the duplex DNA melting point is shown here
analytically by a renormalization group approach.  To achieve this, a
nonperturbative momentum-shell type RG procedure is employed. We
studied the duplex-DNA melting by introducing a rigid chain model,
where the melting is induced by an interfacial term.  A completely
bound two-stranded state at zero temperature is the zero temperature
configuration.  At finite temperatures thermal fluctuations locally
denature the bound state to form bubbles made of two free-chain pairs.
A third similar strand, when added, can again form a duplex with one or
both of the free chains of a bubble. Due to renormalization of
short range interactions close to the duplex melting point an
effective long range three-chain interaction is generated.  The
Efimov DNA is a result of this.  Just as in the quantum Efimov
problem, we show that the Efimov-DNA is associated with a limit cycle
behaviour of the RG flow of the generated three-chain interaction.
Since the interaction parameters for a DNA are easily tunable, by
choosing solvent quality, we hope our results would motivate
experiments in detecting the Efimov effect in polymeric systems.

\appendix

%

\section{Bound State}\label{appen_bound_state}
The Fourier-Laplace transformed bound state partition function is
given by
\begin{eqnarray}
  Z_{\sf b}({\bf k},s)&=&\int d\hat{{\bf n}}\int_{0}^{\infty}dNe^{-Ns}\nonumber\\
 &&\hspace{0.5cm}\times \int d^{3}{\bf r}e^{i{\bf k\cdot r}}\frac{e^{-\epsilon N\Lambda^{2}}}
  {(4\pi)}\delta({\bf r}-N\Lambda\hat{\bf n})\nonumber\\
  &=&\int \frac{d\hat{{\bf n}}}{4\pi}\int_{0}^{\infty}dNe^{-Ns}e^{-\epsilon N\Lambda^{2}}
  e^{iN\Lambda{\bf k\cdot\hat{n}}}\nonumber\\
  &=&\frac{1}{2}\int_{0}^{\pi}\frac{\sin\theta d\theta}{s+\epsilon\Lambda^{2}
    -ik\Lambda\cos\theta}\nonumber\\
  &=&\frac{1}{k\Lambda}\arctan\frac{k\Lambda}{s+\epsilon\Lambda^{2}}
\end{eqnarray}

\section{Rules Of Diagrammatic Calculations}\label{appen_rules}
In this appendix we list all the rules to evaluate diagrams we have
used.

\laplace
Every partition function has two arguments: one space and one length
respectively. With translational invariance, the arguments would be
the difference of the corresponding quantities at the two ends of each
piece.  The dissociation of a bubble or a duplex is our two-body
vertex $g_2$ [Fig. \ref{fig:2}(c) and Fig \ref{fig:3}(b)].  The
interaction between one single chain and a duplex [Fig \ref{fig:3}(c)]
is the three-chain vertex $g_3$. The algebraic expression for any
diagram is obtained by sequentially multiplying the partition
functions and the vertexes as arranged, with integrations over the
intermediate variables. A renormalized vertex is called a vertex
function.

To see how the $k$ conservation appears, consider a  bound state with one bubble
of the type in Fig.  \ref{laplace}.  Applying the above stated rules, this diagram is
evaluated as
\begin{eqnarray}
\label{apndx1}
I({\bf r},N)&=&\int {\sf Z}_{b}({\bf r_1}|z_1){\sf Z}^{2}({\bf r_{1}-r_{2}}|z_{2}-z_{1})\nonumber\\
&&\times {\sf Z}_{b}({\bf r-r_{2}}|N-z_{2})d{\bf r_1}d{\bf r_2}dz_1dz_2.
\end{eqnarray}
The convolution form in the real space leads to a product form in the
Fourier space. We suppress the $z$ integrals for the time being.
Fourier transforming both sides from variable {\bf r} to {\bf k} and
rewriting right hand partition functions in terms of their Fourier
transformed functions we get
\begin{widetext}
\begin{eqnarray}
  \hat{I}({\bf k},N)&=&\int I({\bf r},N)e^{-i{\bf k\cdot r}}d{\bf r}\nonumber\\
  &=&\frac{1}{(2\pi)^{4d}}\int Z_{b}({\bf k_{1}}|z_1)
  Z({\bf k_2}|z_2-z_1)Z({\bf k_3}|z_2-z_1)Z_{b}({\bf k_{4}}|N-z_2)\nonumber\\
  &&\hspace{2cm}\times e^{i{\bf r_1}\cdot({\bf k_1-k_2-k_3})}e^{i{\bf r_2}\cdot({\bf k_2+k_3-k_4})}e^{i{\bf r}\cdot({\bf k_4-k})}
d{\bf r}
  \prod_{j=1,2}\{d{\bf r}_jd z_j\}
\prod_{l=1}^{4}d{\bf k}_l,\nonumber\\
  &=&\frac{1}{(2\pi)^{4d}}\int Z_{b}({\bf k_{1}}|z_1)
  Z({\bf k_2}|z_2-z_1)Z({\bf k_3}|z_2-z_1)Z_{b}({\bf k_{4}}|N-z_2)\times\nonumber\\
  &&\hspace{2cm}\delta({\bf k_1-k_2-k_3})\delta({\bf k_2+k_3-k_4})\delta({\bf k_4-k})
\  dz_1 dz_2  \prod_{j=1}^{4}d{\bf k_j}  .
\end{eqnarray}
 
Performing three $\delta$ function integrals we get the following
relation between different {\bf k}'s:
\begin{eqnarray}
  {\bf k_1=k_2+k_3},\quad{\bf k=k_4},
  \quad {\bf k_4=k_2+k_3}\quad \mbox{and}\hspace{0.25cm} {\bf k=k_1}.
\end{eqnarray}
From these relations it is clear that overall there is one single
{\bf k} and it is conserved at every junction point. Now as there are
four unknown {\bf k}s and we have only three constraints, there is one
undetermined {\bf k} left. This is the characteristic of the loop in
the diagram. Whenever there is a loop there is  an undetermined
{\bf k} over which we have to integrate ($\frac{1}{(2\pi)^d}\int d{\bf
  k}$). The integration over $\bf k$ corresponds to a bubble in real
space with two ends fixed, as, e.g., in $Z^{2}({\bf
  r_{1}-r_{2}}|z_{2}-z_{1})$ in Eq. (\ref{apndx1}).

Now we show how the $s$ conservation appears by evaluating Eq.
(\ref{apndx1}).  Laplace transforming both sides in $N$ and rewriting
every term in the right hand side through their inverse Laplace
transformation we have
\begin{eqnarray}
  A({\bf r},s)&=&\int_{0}^{\infty} e^{-sN}I({\bf r},N)dN\nonumber\\
  &=&\frac{1}{(2\pi i)^{4}}\int_{0}^{\infty}e^{-sN}dN\int
  \prod_{j=1,2} \{d{\bf r}_j dz_j\}\ \prod_{l=1}^{3} ds_l\  Z_{\sf b}
  ({\bf r}_1,s_1)e^{s_1 z_1}
  Z({\bf r}_2-{\bf r}_1,s_2)e^{s_2(z_2-z_1)}\nonumber\\
&&\hspace{4.5cm}
\times Z({\bf r}_2-{\bf r}_1,s_2)
  e^{s_3(z_2-z_1)}
Z_{\sf b}({\bf r}-{\bf r}_2,s_4)e^{s_4(N-z_2)},
\end{eqnarray}
where the $s_{i}$ integrals are the usual Mellin integrals.  We
evaluate $z$ integrations with limit $z_1=0$ to $z_2$, and $z_2=0$ to
$N$ to get
\begin{equation}
  A({\bf{r}},s)=\int_{0}^{\infty}dN \int_{r_i,s_i}(...)\left[\frac{e^{N(s_1-s_4)}-1}{(s_1-s_4)(s_1-s_2-s_3)}
    -\frac{e^{N(s_2+s_3-s_4)}-1}{(s_2+s_3-s_4)(s_1-s_2-s_3)}\right]e^{(s_4-s)N}.
\end{equation}
\end{widetext}
If we now do the $s_i$ integrations using the method of residues,
contributions come only from the poles at $s_1=s_4$, $s_1=s_2+s_3$ and
from the $N$ integration $s=s_4$. As the first bound segment is
labeled by $s_1$, two free chains are labeled by $s_2$ and $s_3$ and
the end bound state is labeled by $s_4$, we see the $s$conservation at
every point with an overall $s$. And similar to the case of {\bf k}
conservation, every loop in Laplace space also possesses one
undetermined $s$ as there are three relations and four $s$'s to be
determined.  So whenever a loop comes, we have to integrate over that
undetermined $s$ ($\frac{1}{2\pi i}\int ds$).

We can now label the diagram in the Fourier-Laplace space using the
above conservation rules as shown in Fig. \ref{laplace}(c). When
evaluated algebraically it gives $Z_{b}^{2}({\bf k},s)I_{0}$ with the
loop integral
\begin{eqnarray}
  \label{eq:142}
   I_{0}&=&\int\frac{d\bar{s}}{2\pi i} \frac{d{\bf q}}{(2\pi)^3} Z\left(\frac{{\bf k}}{2}
  -{\bf q},\bar{s}\right)Z\left(\frac{{\bf k}}{2}+{\bf q},s-\bar{s}\right)\nonumber\\
  &=&\int\frac{d\bar{s}}{2\pi i} \frac{d{\bf q}}{(2\pi)^3}
  \ \frac{1}{\bar{s}-\Lambda^2\ln\mu+\frac{({\bf k}/2-{\bf q})^2}{2}}\nonumber\\
&&\hspace{1cm}  \times\frac{1}{s-\bar{s}-\Lambda^2\ln\mu+\frac{({\bf k}/2+{\bf q})^2}{2}}.
\end{eqnarray}
We evaluate the $\bar{s}$ integral by employing the method of residues.
There is a simple pole at $\bar{s}=\Lambda^2\ln\mu-\frac{({\bf
    k}/2-{\bf q})^2}{2}$. All the contribution to the integral comes
only from this simple pole. So replace the rest of the $\bar{s}$ by its
value at the pole and the prefactor $\frac{1}{2\pi i}$ cancels out
yielding
\begin{eqnarray}
  \label{eq:141}
  I_{0}&=&\int\frac{d{\bf q}}{(2\pi)^3}
  \frac{1}{s{'}+\frac{k^2}{4}+q^2}\nonumber\\
  &=&\frac{4\pi}{(2\pi)^3}\left[\Lambda-\sqrt{s'+k^{2}/4}\arctan\frac{\Lambda}
    {\sqrt{s'+k^{2}/4}}\right]\nonumber
\end{eqnarray}
where $s{'}=s-2\Lambda^2\ln\mu$. A similar kind of integrals also appear
while evaluating three-chain diagrams. We employ this same procedure
to evaluate them.

All the diagrams of this paper are evaluated by using the rules and
procedures discussed in this appendix.

\section{A simple $\beta$ function for $\hat{g}_3$}
\label{sec:limit-cycle-simple}

We propose an extrapolation formula that connects smoothly the flows
for the $\hat{g}_2=0$ case to the $\hat{g}_2=\hat{g}^*_2$ flows.  This
is a toy example to amplify the limit cycle behaviour in a
three-dimensional parameter space, namely, $\hat{g}_2$ and the real
and imaginary parts of $\hat{g}_3$.  For simplicity we choose,
$s_0=1$.

\begin{figure}
\includegraphics[width=\linewidth]{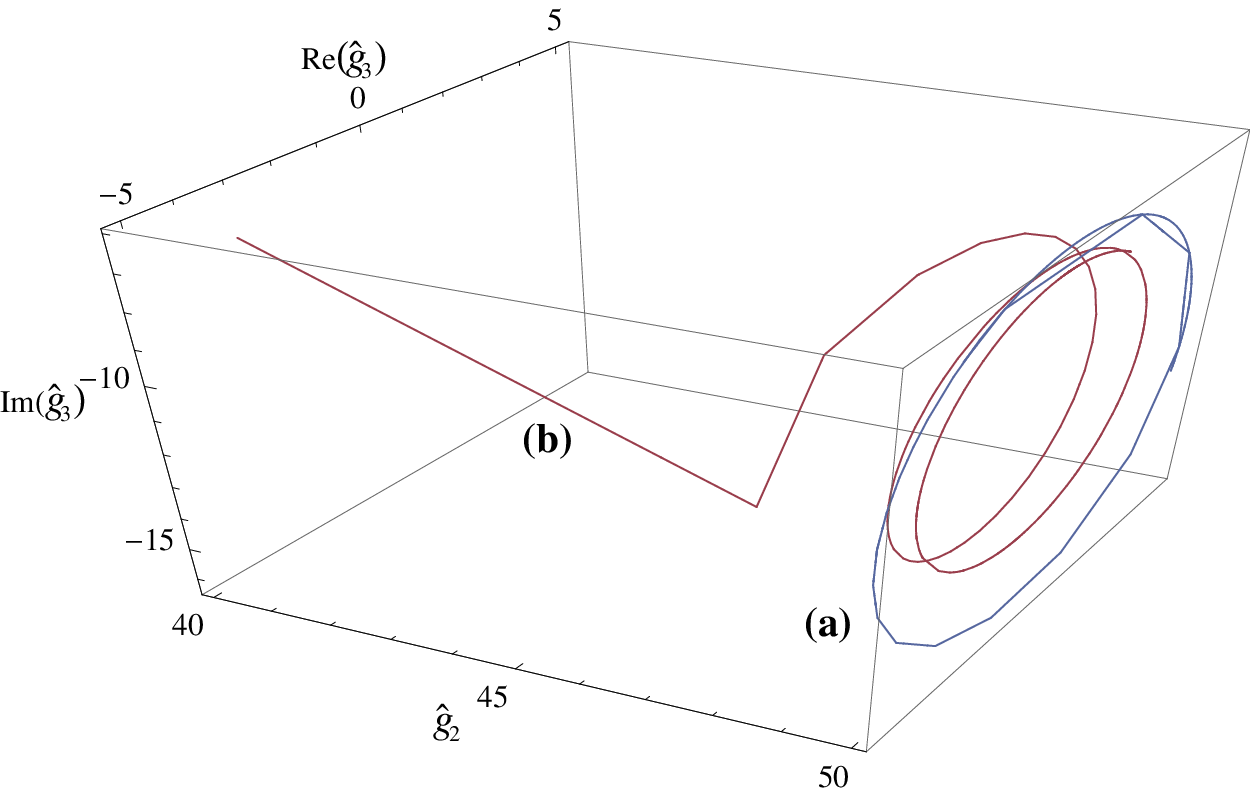}
\caption{(Color online) A schematic diagram of the flow of the RG
  equation, Eq. (\ref{eq:46}), showing the approach to the limit
  cycle. The limit cycle is an ellipse in the $\hat{g}_2=0.5$ plane
  [blue line marked (a)]. A point with off-critical $\hat{g}_2$ is
  shown to approach the planar limit cycle as $\Lambda\to\infty$ [red
  line marked (b)].  }
\label{fig:11}

\end{figure}

Take 
\begin{equation}
  \label{eq:45}
\beta(\hat{g}_2)=\frac{4-d}{2} \hat{g}_2 -\hat{g}_2^2 ,
\end{equation}
with an  unstable fixed point at $\hat{g}_2=0$ and a stable fixed point at
$\hat{g}_2=1/2$.

Use this to write for $d=3$ [Eq.\eqref{eq:39}]
\begin{eqnarray}
  \label{eq:46}
  \beta(\hat{g}_3,\hat{g}_2)&=&  \hat{g}_3  - 2\hat{g}_3 \hat{g}_2  \nonumber\\
 &&  - 4 \hat{g}_2^2 \left ( 2 H - F_2(\hat{g}_2) H^2 - 2 H -1\right),
\end{eqnarray}
where, with $H=-\hat{g}_3/(4\hat{g}_2^2)$ as before, the $2H$ term is
made explicit for comparison with Eq.  (\ref{eq:36}), and we defined
$F_2(\hat{g}_2)=4 \hat{g}_2^2$.  With these choices, we recover both
the limit of $\hat{g}_2=0$ and $\hat{g}_2=1/2$ the critical melting
point.

The flow diagram in the three-dimensional space of $\hat{g}_2,
{\rm{Re}}( \hat{g}_3), {\rm{Im}}( \hat{g}_3)$ shows the approach to
the planar cycle at the melting point.  This is shown in Fig.
\ref{fig:11}. So long as the flow is controlled by the real fixed
points, we see a monotonic flow.  As $\hat{g}_2$ changes, the complex
fixed points take over and we get the loops.

\vfill


\begin{thebibliography}{99}
\bibitem{comm3} If two random walkers or directed polymers starting at
  one spatial point rejoin at another point (including the starting
  point) then it is called a reunion.  The corresponding partition
  function of the bubble decays as a power law in the large length
  limit, $Z_{R}\sim N^{-\psi_{R}}$. This exponent $\psi_{R}$ is called
  the reunion exponent, which plays an important role in determining
  the order of the melting transitions.
\bibitem{comm0}  Although it seems that DNA is a one-dimensional
  object, actually it is not.  The dimensionality $d$ of the embedding
  space in which DNA, and its monomers, live is important in
  determining physical properties of DNA. This dimension specific to
  the model of concern can also be changed, e.g., for a DNA on a
  surface ($d=2$) or a DNA in a solution ($d=3$).  So the theorem that
  there cannot be any phase transition in one dimension is not
  applicable.
\bibitem{commg3} The three-chain coupling $g_3$ here is the negative of
  the same parameter in Ref.  \cite{pal}.  This is done for easy
  comparison with polymer formulations of Refs.  \cite{smbjj,sm1,sm2}.
\bibitem{comm1} For three different chains, if we label the interface
  weight as $g_{2}^{ij}$ for pair $i,j$, then the strand exchange
  would be like $g_{2}^{12}g_{2}^{23} + g_{2}^{13}g_{2}^{23}$. As we
  are considering all the interface weights as $g_2$, the factor of
  $2$ comes in.
\bibitem{comm2} This condition is similar to the ``on-shell'' condition
  used in field theory.
In field theoretical calculations on-shell and off-shell conditions
referred to the center-of-mass frame or non-center-of-mass frame
respectively. Our notion of a relaxed DNA segment is similar to the
on-shell condition where only the Gaussian part ($\sim q^{2}$) is
important. For off-shell considerations one needs to consider
non-Gaussian terms (higher order terms in $q$ ) which we ignore in
this paper.
\bibitem{comm4} Other known examples of this type are
nonrelativistic solid state phyics problems providing examples of
relativistic quantum mechanics, liquid crystals as testing ground of
cosmological theories; see Ref.  \cite{pal}.
\end{thebibliography}
\end{document}